\documentclass[aps,pre,twocolumn,showpacs]{revtex4}

\usepackage{amsmath2000,bm,graphics}

\newcommand{\be}{\begin{equation}}
\newcommand{\ee}{\end{equation}}
\newcommand{\bea}{\begin{eqnarray}}
\newcommand{\eea}{\end{eqnarray}}
\renewcommand{\d}{\text{d}}
\newcommand{\h}{\text{h}}
\newcommand{\braket}[2]{\langle{#1}|{#2}\rangle}
\newcommand{\opket}[2]{{#1} |{#2}\rangle}
\newcommand{\braopket}[3]{\langle{#1}| {#2} |{#3}\rangle}
\newcommand{\ket}[1]{|{#1}\rangle}

\newcommand{\parl}{\parallel}
\newcommand{\Tr}{\text{Tr}}
\newcommand{\wig}[1]{{#1}_{\text{W}}}

\begin{document}


\title{Semiclassical trace formulas for noninteracting identical particles}

\author{Jamal Sakhr}
\email[Correspondence: ]{sakhr@physics.mcmaster.ca}
\author{Niall D. Whelan}
\affiliation{Department of Physics and Astronomy, McMaster University,
Hamilton, Ontario, Canada L8S~4M1}

\date{\today}

\begin{abstract}
We extend the Gutzwiller trace formula to systems 
of noninteracting identical particles. The
standard relation for isolated orbits does not apply since 
the energy of each particle is separately conserved causing 
the periodic orbits to occur in continuous families. The identical
nature of the particles also introduces discrete permutational 
symmetries. We exploit the formalism of Creagh and Littlejohn 
[Phys. Rev. A \textbf{44}, 836 (1991)], 
who have studied semiclassical dynamics 
in the presence of continuous symmetries, to derive  
many-body trace formulas for the full and 
symmetry-reduced densities of states. Numerical studies of the
three-particle cardioid billiard are used to explicitly illustrate and test
the results of the theory.
\end{abstract}

\pacs{05.45.Mt, 03.65.Sq, 05.45.Jn, 05.45.-a.}


\maketitle

\section{Introduction}

In the semiclassical limit of quantum mechanics, the periodic orbits
of the corresponding classical system play a special role in
determining the spectral properties of the quantum system. This
fundamental fact has been a dominant theme in modern semiclassical 
physics and 
was pioneered by Gutzwiller \cite{gutz}, Balian and Bloch
\cite{balbloch}, Strutinsky and Magner \cite{Strut1} and Berry and
Tabor \cite{Tabor}. One of the central results which emerged from this
work is the representation of the density of states
in terms of classical periodic orbits. Such representations are
referred to as trace formulas. Semiclassical
analysis based on the use of trace formulas is now common in many
areas of physics \cite{focusonPOs, bb, resource}. Besides providing a
natural framework for studying the quantum manifestations of classical
chaos \cite{focusonPOs,chaos,dasbuch}, such
analysis has been used in the study of nuclei \cite{Strut1, Strut2,
nuk}, atoms \cite{atoms, anisotropic}, metal clusters
\cite{clust,clust2}, molecules \cite{mole}, chemical systems
\cite{chem}, spins \cite{spins}, Casimir effects \cite{casimir} and
tunneling \cite{tunnel}.   
Trace formulas have also become a prominent analytical tool in the study of
mesoscopic systems \cite{meso}.
New directions continue to be explored \cite{super}. 

Despite the vast utility of trace formulas, their use in the few-body
or many-body context has received little attention.  
Although trace formulas are applicable to interacting
many-body systems, more effort has gone into developing 
semiclassical descriptions of single-particle dynamics in an
appropriate mean field.  One impressive exception is the 
application of the Gutzwiller trace formula to the study of
two-electron atoms and related three-body systems
\cite{2eatoms,threebody}. 
The main difficulty of applying the theory is that periodic orbits
must be found for the interacting many-body system. One approach to
this problem has been proposed in Ref.~\cite{selig} which develops a
particle number expansion of the trace formula. 

In this paper, we consider the extension of the Gutzwiller trace formula to
systems of noninteracting identical particles. 
The semiclassical analysis of such 
systems is quite subtle. As we shall discuss below, if the
particles are noninteracting, there is a continuous symmetry. (Discrete
symmetries in semiclassical trace formulas are discussed in 
Refs.~\cite{discrete,robbins,discrete2} and continuous symmetries in 
Refs.~\cite{Strut1,stephen,nonabelian}.) This symmetry has a profound
consequence; although an $N$-particle system in $d$ dimensions can be
thought of as a single-particle system in $Nd$ dimensions, one cannot
simply apply the Gutzwiller trace formula since the presence of a 
continuous symmetry 
implies the periodic orbits of the full phase
space are not isolated, but rather occur in continuous families.

Recently, we presented 
a semiclassical formalism for the density of states of two
noninteracting identical particles based on an asymptotic analysis of
convolution integrals that arise in the decomposition of the 
semiclassical two-particle density of states \cite{paper1}. In
principle, this approach can be generalised to more than two identical
particles.  In this paper, we pursue a different approach which uses 
the full phase space rather than the individual phase spaces of
each particle. We show that the formalism for continuous symmetries 
\cite{stephen} can be
used to find many-body trace formulas. This 
approach recovers our previous results, but can also be
more easily generalised to arbitrary particle numbers. In addition, it is
conceptually cleaner than the convolution method since spurious
end-point contributions from convolution integrals  do
not arise and therefore need not be explained away. The most important
difference is that the convolution method cannot be used when there
are interactions between the particles whereas the analysis of this
paper can be extended to include interactions \cite{usfut}. 

It is also important to understand the effect of particle symmetry on 
the semiclassical structure of many-body trace formulas. For 
noninteracting identical particles, there are coexisting discrete and
continuous symmetries. While Ref.~\cite{weiden} considers the
symmetry-reduced trace formula due to the discrete permutational
symmetry, it is assumed the periodic orbits are isolated, which is
only true if the particles are strongly interacting (although there is
a brief discussion of the noninteracting case). We include the 
appropriate continuous symmetries to determine the trace formulas 
for the bosonic and fermionic densities of states.    

This paper is organised as follows. In section
\ref{fullphasespace}, we study the case of two noninteracting
identical particles. We first provide the necessary background material
in sections \ref{revqn2pp}-\ref{revsmcl} and then give the 
semiclassical formulation in the full phase space. 
Section \ref{lotsa} considers the extension to $N$
identical noninteracting particles. The symmetry decomposition of the
$N$-particle density of states is examined in section \ref{symmN}. The 
results of a numerical study of the three-particle cardioid billiard 
are then presented to illustrate and test the results of the paper.
We finish the paper with a conclusion and several appendices.

\section{Two Noninteracting Identical Particles}\label{fullphasespace}

\subsection{Quantum density of states}\label{revqn2pp}

The quantum Hamiltonian for two identical noninteracting particles, 
$a$ and $b$, is
\be \label{twoham}
\hat{H} = \hat{h}(\hat{z}_a) + \hat{h}(\hat{z}_b),
\ee
where $\hat{z}_{a/b}$ denote the set of operators
$(\hat{x}_{a/b},\hat{p}_{a/b})$ and $\hat{h}$ is a one-particle
Hamiltonian. The full Hamiltonian (\ref{twoham}) is invariant under the
unitary transformation $\hat{U}$ which exchanges $a$ and
$b$. We define the single-particle energies and eigenstates by
\be
\opket{\hat{h}}{j} = \varepsilon_j\ket{j}.
\ee
Then, the two-particle energies and eigenstates are
$E_{ij}=\varepsilon_i+\varepsilon_j$ and $\ket{i j}$ so that
\be
\opket{\hat{H}}{i j} = E_{ij}\ket{i j}.
\ee
Accordingly, the one and two-particle densities of states are
\bea \label{1pand2pqndos}
\rho_1(\varepsilon) & = & \sum_j    \delta(\varepsilon-\varepsilon_j), \nonumber \\
\rho_2(E) & = & \sum_{i,j} \delta(E-E_{ij}),
\eea
and they are related by the convolution identity
\be \label{2ppconv}
\rho_2(E) = \left(\rho_1*\rho_1\right)(E).
\ee
A useful result is the relation between the density of states
and the trace of the energy Green function or resolvent. We define
$g(E)={\Tr}(\hat{G}(E))$, where $\hat{G}(E) = {1/(
E-\hat{H})}$ is the one-sided Fourier transform of the quantum
propagator. In terms of the resolvent,
\be
\rho(E) = -{1 \over \pi} \text{Im} \left \{ g(E + i \epsilon) \right \},
\ee
and this applies for either the one or two-particle density of states as
long as we use the appropriate resolvent on the right-hand side. In
the limit $\epsilon \rightarrow 0^+$, the exact density of states is
recovered \cite{bb}. Henceforth, the $i \epsilon$ will be implicit.

\subsection{Symmetry decomposition}

The most interesting aspect of the existence of identical particles is
the fact that only certain states are occupied, the fully symmetric
ones if the particles are bosons or the fully antisymmetric ones if
the particles are fermions. It is important to understand how the
above discussion decomposes when we consider the separate densities of
symmetric and antisymmetric states. Although not absolutely necessary for the 
present discussion, it will be useful for later to introduce projection 
operators. As mentioned, the Hamiltonian
(\ref{twoham}) is invariant under exchange of the particles $a$
and $b$, an operation we denote by $\varsigma$ (leaving the particles
unchanged we denote by $\iota$). There is a two-element discrete group
comprised of these operations and the representation of these group 
elements in the Hilbert space (\textit{i.e.} the quantum operators 
that exchange the 
particles) are $\hat{U}$ and $\hat{I}$. Both of these operators
commute with $\hat{H}$. This is a simple group with two irreducible
representations (irreps) which we identify as the bosonic (symmetric)
representation and the fermionic (antisymmetric) representation. Given
an arbitrary state with components belonging to both irreps,
we can project out the portion belonging to each irrep 
through the use of the projection operators
\cite{hamermesh}
\be \label{projector}
\hat{P}_\pm = {1\over 2}\left(\hat{I} \pm \hat{U}\right),
\ee
where the $\pm$ refer to the bosonic and fermionic irreps,
respectively.

In terms of these projection operators, the bosonic and fermionic
densities of states are given as
\be
\rho_\pm(E) = {\Tr} \left(\hat{P}_\pm\delta(E-\hat{H})\right).
\ee
The sum of the bosonic and fermionic densities is the
complete two-particle density of states, $\rho_2(E)$. The difference
is given by ${\Tr}(\hat{U}\delta(E-\hat{H}))$ and
expressing the trace in the energy eigenbasis, 
\bea \label{sym2pden}
\rho_+(E)-\rho_-(E) 
& = &\sum_{i,j}\braopket{i j}{\hat{U}\delta(E-\hat{H})}{i j} \nonumber \\
& = &\sum_{i,j}\braket{j i}{i j}\delta(E-E_{ij}) \nonumber \\
& = &\sum_j\delta(E-2\varepsilon_j),
\eea 
where we have used the fact that $\hat{U}$ exchanges the state labels
in the second line and the fact that $E_{jj}=2\varepsilon_j$ in the third. The
final line we recognise as $\rho_1(E/2)/2$ and thereby conclude
\be \label{symmreddos}
\rho_\pm(E) = {1\over 2}\left(\rho_2(E)\pm{1\over
2}\rho_1\left({E\over 2}\right)\right).
\ee
\subsection{Review of semiclassical formulation}\label{revsmcl}

It is common to decompose the semiclassical density of states into smooth and
oscillatory components. For the one-particle density
\be \label{decomp}
\rho^{\text{sc}}_1(\varepsilon) = \bar{\rho}_1(\varepsilon) + 
\tilde{\rho}_1(\varepsilon),
\ee
where $\bar{\rho}$ and $\tilde{\rho}$ denote the smooth and oscillating
components, respectively. There is an extensive literature on this
decomposition \cite{bb}. We adopt the point of view that
one can simply use the leading-order contribution of each component.
We do not consider the subtle issues related to
the asymptotic nature of this decomposition (see, for example, 
Refs.~\cite{howls, howls2, spurious}). 
For analytic potentials 
in $n$ dimensions, the leading-order term for the smooth density of states is
\be \label{smooth}
\bar{\rho}_1(\varepsilon) \approx {1\over (2\pi\hbar)^n}\int \d z 
\delta(\varepsilon-h(z)),
\ee
where $z$ collectively denotes the $2n$ classical phase space
coordinates and $h(z)$ is the classical Hamiltonian (for an exception
to this general result, see Refs.~\cite{x2y2}). There are corrections to
(\ref{smooth}) involving derivatives of the delta function in the
integrand. The first correction is $O(\hbar^2)$. For a two-dimensional
billiard, the analogous expression is
\be \label{smooth_bill}
\bar{\rho}_1(\varepsilon) \approx {\alpha {\mathcal{A}} \over 4\pi} \pm 
   {\sqrt{\alpha} {\mathcal{L}} \over 8\pi\sqrt{\varepsilon}} 
 + {\mathcal{K}} \delta (\varepsilon),
\ee
where $\alpha=2m/\hbar^2$, ${\mathcal{A}}$ and ${\mathcal{L}}$ 
refer to the area 
and perimeter, respectively and the $\pm$ refers to Neumann and
Dirichlet boundary conditions, respectively. The third term 
\be \label{curvterm}
{\mathcal{K}} = {1 \over 12\pi} \oint\d l \kappa(l) + {1 \over 24\pi} 
\sum_i {{\pi^2 - \theta_i^2} \over \theta_i}
\ee
is the average curvature integrated along the boundary with
corrections due to corners with angles $\theta_i$. It does not actually 
contribute to the density of states, but rather to its first integral and this
term will be used in section \ref{num3pp}. 
There are also corrections involving 
powers and derivatives of the curvature (see 
Refs.~\cite{howls,howls2} for more exhaustive studies). Similar results hold 
for higher dimensional billiards (see Ref.~\cite{bb}).

The oscillating component can be written as \cite{gutz}
\be \label{rough}
\tilde{\rho}_1(\varepsilon) \approx -{1\over\pi}\text{Im}\left\{
\sum_\gamma A_\gamma(\varepsilon) \exp i \left({S_\gamma(\varepsilon) 
\over\hbar} -\sigma_\gamma{\pi\over2}\right)\right\}, 
\ee
where $\gamma$ labels the periodic orbits of the system,
$S_\gamma$ is the classical action integral along the orbit 
and $\sigma_\gamma$ is a topological index \cite{index}
counting the caustics in phase space encountered by the
orbit. $A_\gamma$ is the amplitude of the periodic orbit and depends
on the specific nature of the system, for example, whether the orbit is 
isolated or not and if so on its stability. For the case of isolated
periodic orbits,
\be \label{isolated}
A_\gamma(\varepsilon) = {1 \over i\hbar}
{T_\gamma^0(\varepsilon) \over \sqrt{\left|{\det}(\tilde{M}_\gamma-I)\right|}},
\ee
where $T_\gamma^0$ is the primitive period of the orbit and 
$\tilde{M}_\gamma$ is the $2(n-1)\times 2(n-1)$ symplectic stability
matrix on any Poincar\'e section to which the orbit is transverse. Its
eigenvalues give the stability exponents of the orbit.

As stated above, the density of
states for two noninteracting particles is the 
autoconvolution of the one-particle density of states (\ref{2ppconv}). 
Formally, the semiclassical two-particle density of states is the 
autoconvolution of (\ref{decomp}), that is,  
\be \label{f2p}
\rho^{\text{sc}}_2(E) = \left(\rho^{\text{sc}}_1 * \rho^{\text{sc}}_1\right)
(E) = \bar{\rho}_2(E) + \tilde{\rho}_2(E),
\ee
where
\begin{subequations}\label{smcldos2ppconv}
\bea
& \bar{\rho}_2(E)    =  \left(\bar{\rho}_1*\bar{\rho}_1\right)(E), 
\label{sp2p} \\
& \tilde{\rho}_2(E)  =  2\left(\bar{\rho}_1*\tilde{\rho}_1\right)(E) + 
\left(\tilde{\rho}_1*\tilde{\rho}_1\right)(E). \label{op2p}
\eea
\end{subequations}
The mixed term $2\left(\bar{\rho}_1*\tilde{\rho}_1\right)
(E)$ also belongs to the oscillating component of $\rho^{\text{sc}}_2(E)$. 
This is because an asymptotic endpoint analysis of the convolution
integral results in an oscillatory function as shown in Ref.~\cite{paper1}
where all components have been evaluated and given explicit semiclassical
interpretations in terms of one and two-particle dynamics which
support this decomposition. 
We also showed above how the difference between the bosonic and fermionic
densities is given by the one-particle density of states. Formally, we may 
write (analogous to Eq.~(\ref{symmreddos})), 
\be \label{symmreddossmcl}
\rho^{\text{sc}}_\pm(E) = {1\over 2}\left(\rho^{\text{sc}}_2(E)\pm{1\over
2}\rho^{\text{sc}}_1\left({E\over 2}\right)\right).
\ee
The formal results (\ref{f2p}-\ref{symmreddossmcl}) can also  
be understood from a semiclassical analysis in the full phase space. 
This analysis is not only more fundamental, 
it is necessary if one wants to include 
interparticle interactions since the particle dynamics then 
become coupled and we can no longer make use of calculations 
that involve the individual one-particle phase spaces.
In the following sections, we derive trace formulas for the 
full and symmetry-reduced densities of states 
from semiclassical calculations in the full two-particle phase
space. 
Since we are mainly concerned with the extension of the Gutzwiller
theory, we focus on the fluctuating part of the density of
states. However, since the smooth part is important in constructing
the complete density of states, we have provided a discussion of the
two-particle Thomas-Fermi term (and the associated symmetry
decomposition based on the theory of Ref.~\cite{pottf}) in 
Appendix \ref{TF2pcal}.
To calculate the fluctuating part of the density of
states, we need to find all periodic orbits in the full phase space at
a specified energy $E$. 

\subsection{Two-particle dynamics in the full phase space}\label{dyn2ppfps}

The two identical particles, $a$ and $b$, evolve independently in their own
one-particle configuration space which we denote as having dimension
$d$ so that the one-particle phase spaces are of dimension $2d$. The
full two-particle configuration space has dimension $2d$ and the
corresponding phase space is of dimension $4d$.  We reserve the symbol
$z$ to collectively denote these $4d$ phase space coordinates and will
use $z=(z_a,z_b)$ where $z_{a/b}$ denote the $2d$-dimensional
one-particle phase space coordinates of each particle. 
We recall that dynamics in the full phase
space consists of each particle evolving separately in its own phase
space. The dynamics of $z$ are defined through one-particle dynamics
by $\Phi_t z = (\phi_t z_a,\phi_t z_b)$, where $\phi_t$ is the flow
for one particle. The
(noninteracting) two-particle Hamiltonian is $H(z)=h(z_a)+h(z_b)$, 
where $h(z_{a/b})$ is a one-particle Hamiltonian. 

We seek periodic orbits with phase space coordinates $z'$ such that
$\Phi_T z' = z'$ for some period $T$. This is possible if the two
particles are on (generally distinct) periodic orbits
with the same period. In general, two arbitrary periodic orbits will
have different periods. However, there is a parameter which we can
vary, namely the way in which the total energy is partitioned between
the two particles. Generally, we can find an energy $E_a$ (and
$E_b=E-E_a$) such that the two periods are the same. We will assume
henceforth that there is only one energy $E_a$ for which there is a
solution. (This assumption can be relaxed at the cost of heavier
notation.) There is another way to have a periodic orbit in the full
phase space; one particle can evolve dynamically on a periodic orbit 
with all of the energy while the other is stationary at a fixed point of the
potential. This is discussed later. 

\begin{figure}
\scalebox{0.355}{\includegraphics*{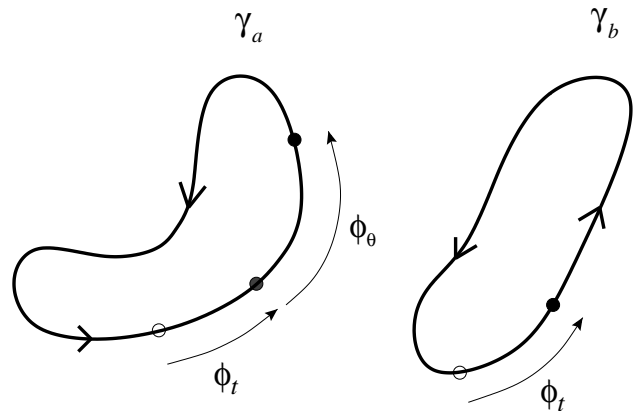}}
\caption{\label{fPOeq2POs} Two periodic orbits $\gamma_a$ and
$\gamma_b$ which constitute a periodic orbit
$\Gamma$ of the full phase space. The full Hamiltonian 
$H(z_a,z_b)$ generates time
translations for both particles (as denoted by the single-particle
flow $\phi_t$ acting on both particles) while the single-particle Hamiltonian
$J=h(z_a)$ generates time translations for particle $a$ while leaving
particle $b$ fixed (as denoted by the single-particle flow
$\phi_{\theta}$ acting on particle $a$ only). The flows generated by $H$ and
$J$ are $\Phi_t$ and $\Psi_\theta$, respectively. A combination of such
flows (cf.  Eq.~(\ref{flowscommute})) is shown here.}
\end{figure}

\subsubsection{Dynamical periodic orbits}

If \emph{both} particles are on periodic orbits, we call the full phase
space periodic orbit a \textit{dynamical} periodic orbit. We first note
that such 
orbits occur in continuous families. To see this, imagine that a full
phase space periodic orbit consists of one particle on a 
periodic orbit $\gamma_a$ and the other particle on a distinct
periodic orbit $\gamma_b$ (see Fig.~\ref{fPOeq2POs}) and
that the energy partition is such that both orbits have the same period
$T$. We have complete freedom in specifying which points on the
respective orbits we choose as initial conditions. Given that we
define $t=0$ to be when particle $b$ is at some specified point on
$\gamma_b$, we can vary the position of particle $a$ on $\gamma_a$. By
changing its initial position along the orbit, we map out a continuous
family of congruent periodic orbits.

This can be formalised as follows. We note that in addition to the
total Hamiltonian $H$, there is a second constant of motion $J=h(z_a)$
in involution with $H$. It generates time translations of particle $a$ while
leaving particle $b$ fixed. (In fact, $J$ can be chosen as any
linear combination of $h(z_a)$ and $h(z_b)$ as long as it is
independent of $H$.) 
Flows generated by $J$ are denoted by
$\Psi_\theta$ and are mapped in the full phase space as follows: $\Psi_\theta
z=(\phi_\theta z_a,z_b)$. The symmetry parameter $\theta$ is conjugate
to $J$ and has the dimension and interpretation of time. However,
since it only measures the evolution of particle $a$, 
it is not time in the usual sense and we will follow
the notation of Ref.~\cite{stephen} in denoting the paramater by $\theta$. 
A combination of flows in $H$ and $J$ is
\be \label{flowscommute}
\Phi_t \Psi_\theta z =(\phi_t \phi_{\theta}z_a,\phi_t z_b)=
(\phi_{t+\theta}z_a,\phi_t z_b). 
\ee
Since $\Psi_\theta$ and $\Phi_t$ commute and separately conserve both
$H$ and $J$, the surface mapped out by these flows has constant $H$
and $J$ (\textit{i.e.} $H=E$ and $J=E_a$) . Starting at some point on the
full phase space periodic orbit, flows in $H$ and $J$ map out a
two-dimensional torus. This means there is a 1-parameter degenerate
family of periodic orbits (the other dimension is parameterised by
time and is present even in the case of isolated orbits). Therefore,
we cannot use the Gutzwiller trace formula for isolated orbits since
it will give a spurious infinity. Due to the continuous
family, there is one fewer stationary phase integrals to be done in evaluating
the trace so that this family of orbits contributes
$O(1/\sqrt{\hbar})$ more strongly than an isolated orbit and the
calculation of its amplitude must be performed carefully.

For the present, we assume that there are no symmetries other than $J$
so that all periodic orbits of the one-particle phase space are isolated. The
flow directions generated by $H$ and $J$ are stable as are the two
directions transverse to the constant $H$ and $J$ surfaces. Thus,
there are four directions of neutral stability in phase space. The
remaining $(4d-4)$ directions decompose into separate subspaces of
dimension $(2d-2)$ within each of which there are the standard
symplectic possibilities for stability.

In general, the leading-order contribution of one $f$-parameter 
family of orbits
(generated by Abelian symmetry) to the resolvent is \cite{stephen}
\begin{equation} \label{stephensequation} \begin{split}
\tilde{g}_\Gamma(E)& = {1\over i\hbar} {1\over
\left({2\pi\hbar}\right)^{f/2}} {T^0_\Gamma
V^0_\Gamma \over
\left|{\partial \mathbf{\Theta}\over \partial \mathbf{J}}\right|_\Gamma^{1/2}
\left|{\det}\left(\tilde{M}_\Gamma - I\right)\right|^{1/2}} \\
& \qquad\times \exp{i\left({S_\Gamma(E)\over \hbar}-(\mu-\delta)_\Gamma 
{\pi\over2}-f {\pi\over4}\right)}.
\end{split} 
\end{equation}
This contribution is $O(1/{\hbar}^{f/2})$ stronger than an isolated
periodic orbit. As mentioned above, every constant of motion implies
one fewer stationary phase integrals and therefore $f$ fewer powers of
$\sqrt{\hbar}$ in the prefactor. For a similar reason, there is an
additional phase factor of $-f \pi/4$. The total contribution to the
resolvent is a sum over all families of periodic orbits, the capital
$\Gamma$ indicating that these are indeed families and not isolated
orbits as in the more familiar Gutzwiller trace formula. In our case,
the sum over $\Gamma$ can be expressed as a double sum over $\gamma_a$
and $\gamma_b$ indicating the periodic orbits on which
the particles are evolving.  We now describe the various factors in
(\ref{stephensequation}) and explain what they are in the present
situation for which $f=1$.

The volume term, $T^0_\Gamma V^0_\Gamma$ is the integral over the
flows generated by $H$ and $J$, $\oint_\Gamma \d t
\d\theta$, integrated over the periodic orbit family. The time
integral gives the period of the family $T_\Gamma=
T_{\gamma_a}(E_a)=T_{\gamma_b}(E_b=E-E_a) \equiv T$ while the $\theta$
integral gives $V_\Gamma=T_{\gamma_a}(E_a)$ since a flow in $J$ by
that amount returns particle $a$ to where it began. (Hence, the intial phase
space coordinate is mapped back to itself under the dynamics.) 
However, there can be discrete symmetries such that a
combination of flows in $H$ and $J$ for less than $T$ restores the intial 
conditions. This situation occurs when
one or both particles is on a repetition of some primitive periodic
orbit. To see this, suppose that particle $a$ is on the $n_a$'th
repetition of its orbit while particle $b$ is on the $n_b$'th
repetition of its orbit. Then, the torus is partitioned into $n_an_b$
equivalent segments and the \textit{primitive} volume term is $T^0_\Gamma
V^0_\Gamma=T_\Gamma V_\Gamma/n_an_b$. However, the full periods 
are defined through the primitive periods by 
$T_{\gamma_a}(E_a)=n_aT^0_{\gamma_a}(E_a)$ and similarly for particle
$b$. Thus, $T^0_\Gamma
V^0_\Gamma=T^0_{\gamma_a}(E_a)T^0_{\gamma_b} (E_b=E-E_a)$, which is
the product of the primitive periods.

$\tilde{M}_\Gamma$ is a $(4d-4)\times (4d-4)$ matrix linearising
motion on a reduced surface of section. Specifically, it is the
section at constant $(H,J,x_{\parl a},x_{\parl b})$ where 
$x_{\parl {a/b}}$ are chosen so that the dynamics are transverse to
the surface on which they are both constant. In our case, this section
is simply the direct product of the normal
Poincar\'e surfaces of section for each of the two motions (where one
would specify the one-particle energy and some fixed coordinate in
each case). As a result, $\tilde{M}_\Gamma$ has a block diagonal
structure since there is no coupling between the two particle
spaces. We conclude that 
${\det}(\tilde{M}_\Gamma - I)={\det}(\tilde{M}_{\gamma_a} - I){\det}
(\tilde{M}_{\gamma_b} - I)$, 
where $\tilde{M}_{\gamma_a/\gamma_b}$ are the stability matrices of
the separate one-particle periodic orbits and $I$ is the appropriately
dimensioned unit matrix on both sides of the equality.

The anholonomy term 
$\left({\partial \Theta / \partial J}\right)_\Gamma$ measures the
amount by which orbits that are periodic in the symmetry-reduced
dynamics fail to be periodic in the full phase space.  Suppose we vary
the value of $J$ infinitesmally while keeping the total energy fixed;
in our case this amounts to a slight change of the energy partition
between the two particles. The periodic orbit is launched as before
with the same initial conditions except for $p_\parl$ (the momentum
conjugate to $x_\parl$) which must be changed appropriately to effect
the change in $J$. After the original period $T$, an initial phase
space coordinate will not be mapped back to where it began, 
but rather infinitesmally close to this initial condition. A
flow in $H$ for some extra amount of time $\Delta t$ and a flow in $J$
by an extra amount
$\Delta \theta$ (or vice-versa since the flows commute)
closes the orbit in the full phase space. The factor ${\partial\Theta /
\partial J}$ is simply the ratio $\Delta\theta/\Delta J$ (in the limit
that $\Delta J\rightarrow 0$). ($\Theta$ is capitalised to stress that
$J$ and $\theta$ can also be used as labels of families of surfaces,
in which case, this factor can be interpreted as a Jacobian for a
change of label from $J$ to $\theta$.) Recall that the value of
$J=h(z_a)$ is just the energy of particle $a$. If $J_a \rightarrow
J_a+\Delta J_a$, then $E_b \rightarrow E_b-\Delta J_a$, since the
total energy is fixed. $\gamma_a$ now has a perturbed period, $T+\Delta
T_{\gamma_a}=T+T'_{\gamma_a}\Delta J_a$ while $\gamma_b$ now has a perturbed
period $T+\Delta T_{\gamma_b}= T-T'_{\gamma_b} \Delta J_a$, where the
primes denote differentiation with respect to energy:
\be
T'_{\gamma_a}= \left. {{\d} T_{\gamma_a}(\varepsilon) \over {\d} 
\varepsilon}\right|_{\varepsilon=E_a}, \quad T'_{\gamma_b}= - \left. 
{{\d} T_{\gamma_b}(E-\varepsilon) \over {\d} 
\varepsilon}\right|_{\varepsilon=E_a}.
\ee
Let $z'=(z'_a,z'_b)$
and $z$ denote the initial and final phase space coordinates, 
respectively. Then, after the original period $T$, 
\be
\Phi_Tz'=z = (\phi_{-\Delta T_{\gamma_a}}z_a',\phi_{-\Delta T_{\gamma_b}}z_b').
\ee
We need to find $(\Delta t,\Delta\theta)$ which map $z$ back to $z'$.
Using Eq.~(\ref{flowscommute}), the condition for a periodic orbit  
$\Phi_{\Delta t}\Psi_{\Delta \theta} z=z'$ implies $\Delta t=\Delta
T_{\gamma_b}$ and $\Delta\theta=\Delta T_{\gamma_a}-\Delta T_{\gamma_b}$ 
so that
\be \label{anholonomy}
{\partial \Theta\over \partial J} = T'_{\gamma_a} + T'_{\gamma_b}.
\ee
The action $S_\Gamma(E)=S_{\gamma_a}(E_a)+S_{\gamma_b}(E_b=E-E_a)$ is
the action of the periodic orbits in the family (all orbits in
$\Gamma$ have the same action because of symmetry). 
Finally, we discuss the phase indices. $\mu_\Gamma$ is determined from
the dynamics in the symmetry-reduced surface of section in the same
way as for isolated orbits in the usual Gutzwiller trace formula and
following the same logic as above, 
$\mu_\Gamma=\sigma_{\gamma_a}+\sigma_{\gamma_b}$.  $\delta_\Gamma$ is
defined as the number of positive eigenvalues of $\left({\partial
\Theta/ \partial J}\right)_\Gamma$ \cite{indices}.  In this case, the
anholonomy term is simply a scalar and therefore $\delta_\Gamma=1$ if
the Jacobian is positive and $\delta_\Gamma=0$ if the Jacobian is
negative. We conclude that the contribution to the resolvent from one
family of dynamical orbits is 
\begin{widetext}
\be \label{finalresult}
\tilde{g}^{\d}_\Gamma(E) =
{i\hbar\over \sqrt{2\pi\hbar}}
\left(
{T^0_{\gamma_a}(E_a)\over i\hbar}
{\exp i \left({S_{\gamma_a}(E_a)\over\hbar}-\sigma_{\gamma_a}{\pi\over2}
\right) \over
\left|{\det}\left(\tilde{M}_{\gamma_a} - I\right)\right|^{1/2}}
\right)
\left(
{T^0_{\gamma_b}(E_b)\over i\hbar}
{\exp i \left({S_{\gamma_b}(E_b)\over\hbar}-\sigma_{\gamma_b}{\pi\over2}
\right) \over
\left|{\det}\left(\tilde{M}_{\gamma_b} - I\right)\right|^{1/2}}
\right)
{\exp i \left(\delta_\Gamma {\pi\over2} -{\pi\over4}\right) 
\over \sqrt{\left|T'_{\gamma_a} + 
T'_{\gamma_b}\right|}}.
\ee
\end{widetext}
As mentioned above, we assumed that there is only one energy partition
such that both
particles have the same period. This will be the case when the period
is a monotonic function of energy, which is a typical situation. If
the period is a more complicated function of energy, there may be
further solutions and if so then one must have a sum over
$(\gamma_a,\gamma_b)$ for each possible solution of this condition, 
but we suppress this possibility for notational simplicity.

We obtained Eq.~(\ref{finalresult}) in Ref.~\cite{paper1} by doing a stationary
phase analysis of the direct autoconvolution of Eq.~(\ref{rough}).
(The phase index $\nu$ in Eq. (18) of Ref.~\cite{paper1} has a
different definition than $\delta_\Gamma$ in Eq. (\ref{finalresult}),
but the overall phase is consistent in the two formulas.) The
condition of stationary phase immediately implied that the energy must
be partitioned so that the periods of the two orbits are
the same. The stationary phase integral then introduces a factor of
$\sqrt{\hbar}$ as well as the sum of the second derivatives of the
actions with respect to energy evaluated at the stationary phase
energy. This is precisely the first derivatives of the periods with
energy. Thus, we have shown how these two different approaches yield
consistent results.

Either particle can execute any number of repetitions of its primitive 
orbit. If particle $a$ executes $l_{\gamma_a}$ repetitions and particle $b$
executes $l_{\gamma_b}$ repetitions, then the energy must be partitioned so
that $l_{\gamma_a}T_{\gamma_a}(E_a) = l_{\gamma_b}T_{\gamma_b}(E_b)$. 
As discussed above, the volume term 
remains unchanged as the square of the primitive periods due
to the discrete symmetry in the family of orbits. The action of this
orbit is $l_{\gamma_a}S_{\gamma_a}(E_a)+l_{\gamma_b}S_{\gamma_b}(E_b)$, 
and similarly for the phase index.
The separate monodromy matrices are raised to the
appropriate power. The anholonomy term follows from the discussion above
Eq.~(\ref{anholonomy}) as $l_{\gamma_a}T'_{\gamma_a}(E_a) +
l_{\gamma_b}T'_{\gamma_b}(E_b)$. 
Assuming that the periods are monotonic in the
energy, the index $\delta_\Gamma$ is unchanged. Otherwise, it 
depends on the sign of $l_{\gamma_a}T'_{\gamma_a} +
l_{\gamma_b}T'_{\gamma_b}$. Therefore, 
the previous expression continues to apply, but the actions, phase
indices and periods are multiplied by the appropriate value
of $l_{\gamma_a / \gamma_b}$ and the separate monodromy matrices are 
raised to the appropriate
power. This dependence can also be understood to be already
contained in the definition of the various orbit properties.

We observe that the amplitude of (\ref{finalresult}) is proportional
to the product of the amplitudes for the single-particle dynamics.
The trace formula for two noninteracting particles contains 
an additional prefactor of $i\hbar/\sqrt{2\pi\hbar}$, a
factor involving the derivatives of the periods with
respect to energy (and the associated phase index $\delta$) and an
additional phase factor of $\pi/4$. This result generalises to cases
where the amplitudes are not given by (\ref{isolated}). We simply
replace the single-particle amplitudes in large brackets by the
equivalent ones for the system under consideration. This can be
understood by noting that the only coupling between the particles is
as we have described and any further symmetry can be handled within
the single-particle phase spaces. This conclusion can also be
understood in the convolution picture by simply using the appropriate
single-particle amplitudes when doing the stationary phase analysis
\cite{paper1}.

\subsubsection{Symmetry decomposition: dynamical pseudo-periodic orbits (DPPOs)}
\label{dynppo}

As discussed initially by Gutzwiller \cite{discrete} and later in more
generality by Robbins \cite{robbins}, in the presence of a discrete
symmetry, the fluctuating density of states can be decomposed among
the various irreducible representations (irreps). 
For the two-particle case, this is
simply the symmetric (bosonic) and antisymmetric (fermionic) cases. To
evaluate the separate densities of states, one must calculate
$g_\pm(E)=\Tr (\hat{P}_\pm\hat{G}(E))$ using the projection
operators in (\ref{projector}). The first term of the projection
operator results in the standard sum over dynamical periodic
orbits (\ref{finalresult}). There is a factor of $1/2$ which indicates
that this contribution is simply divided evenly between the symmetric
and antisymmetric spectra. It is the second term of the projection
operator which requires careful analysis.

The oscillating part of $\Tr (\hat{U}\hat{G}(E))$ can be expressed in
terms of orbits on which particles begin at a point in phase space,
evolve for some time $T$, are then exchanged using the classical
analogue of $\hat{U}$ with the net result that the particles are returned to their
initial conditions. We call these orbits \textit{pseudo-periodic} 
to distinguish them
from the (standard) dynamical periodic orbits discussed earlier. 
We first define the symplectic mapping
$u$ corresponding to classical particle exchange as
$u(z_a,z_b)=(z_b,z_a)$. It has the property that $u^2$ is the identity
mapping. The combination of time evolution for time $t$ and
particle exchange maps a phase space point $z'=(z_a',z_b')$ to
$z=u\Phi_tz'=(\phi_tz_b',\phi_tz_a')$. To find orbits which are
periodic under these combined operations, we require phase space
coordinates $z'$ and periods $T$ such that $z'=u\Phi_Tz'$. Applying
this combined operation twice, we find that $z'=\Phi_{2T}z'$. This is
just the condition for a periodic orbit of period $2T$ in the full
phase space without particle exchange. So we conclude that the initial
coordinate $z'$ must be on a periodic orbit of the full phase space.
However, this condition is still more restrictive since the above
considerations also imply that after time $T$, particle $a$ must be where
particle $b$ began and vice-versa. This is only possible if the two
particles are traversing the \emph{same} periodic orbit, with the same
energy and furthermore are exactly half a period out of phase. We shall call this 
a type-1 dynamical pseudo-periodic orbit (DPPO).  
There is also the degenerate case
where both particles begin and evolve together. This shall be called 
a type-0 DPPO and is discussed below. 

\begin{figure}
\scalebox{0.3}{\includegraphics{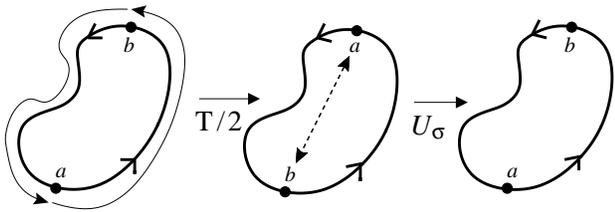}}
\caption{\label{symdecomp2pp} A dynamical pseudo-periodic orbit (DPPO) of 
the full (two-particle) phase space is
constructed by placing two particles on a periodic orbit of the
one-particle phase space. If $E_a=E_b$ and the particles
are half a period out of phase, then after the combined operations of
time evolution and particle exchange, the initial conditions are restored.}
\end{figure}

Therefore, the set of possible pseudo-periodic orbits is much more
restricted than the set of standard periodic orbits since we have only
contributions when both
particles are executing the same dynamics. Furthermore, these orbits
are isolated and do not come in a 1-parameter family. The existence
of families for the standard periodic orbits is due to the freedom in
specifying the relative phases of the two motions. We no longer have
this freedom. This immediately implies that contributions from 
pseudo-periodic orbits
will be weaker by $\sqrt{\hbar}$ because there is one more
stationary phase integral to do than for the standard periodic
orbits. (This can also be understood from the fact that particle
exchange does not conserve the separate energies and so does not
commute with $J$.) Therefore, the usual Gutzwiller trace formula
applies and we use it to determine the actions, periods and
stabilities of these isolated pseudo-periodic orbits.

Consider an arbitrary periodic orbit $\gamma$ of the one-particle
phase space with period $T_\gamma$
and choose some arbitrary initial condition on it which we shall call
$z'_a$. To have a pseudo-periodic orbit in the full phase space, we
can begin at $z'=(z'_a,z'_b=\phi_{T_\gamma/2}z'_a)$. If we flow for a
time $T_\gamma/2$ and then exchange the particles, we map $z'$ onto
itself (see Fig.~\ref{symdecomp2pp}).  Therefore, the set of
pseudo-periodic orbits in the full two-particle phase space is
one-to-one with the set of standard periodic orbits in the
one-particle phase space. The periods of the pseudo-periodic orbits in
the full phase space are one-half of the periods of the corresponding standard
periodic orbits in the one-particle phase space. Nevertheless, when
evaluating the trace integral we must integrate over all initial
conditions on the orbit and this gives a full factor of $T^0_\gamma$ 
in the amplitude. The
actions and phase indices for the pseudo orbit are the same as
for the standard orbit; although we integrate for only half the
time, both particles are in motion and between them, they execute one
full motion of the periodic orbit. The stability matrix in the full
phase space requires careful analysis. Let $\tilde{M}_\gamma$ be the
stability matrix of the full periodic orbit $\gamma$ of the
one-particle phase space and $\tilde{M}_{\gamma'}$ be the stability
matrix of the pseudo-periodic orbit ${\gamma'}$ in the full phase
space. It is shown in Appendix \ref{determinants} that 
$\det(\tilde{M}_{\gamma'}-I) = 4 \det
(\tilde{M}_\gamma-I)$, where on each side of the equation $I$ is
understood to be an appropriately dimensioned unit matrix.
We conclude that the contribution of this orbit to the oscillating part of 
$\Tr (\hat{U}\hat{G}(E))$ is
\be
\tilde{g}^{\text{d}}_{\varsigma}(E)={1\over 2 i \hbar} 
{T^0_\gamma \over \sqrt{\left|{\det}(\tilde{M}_\gamma-I)\right|}} 
\exp i \left({S_\gamma \over \hbar}-\sigma_\gamma {\pi\over2}\right), 
\ee
where all classical quantities are evaluated at the single-particle
energy $E/2$. (Recall the symbol $\varsigma$ denotes the group 
element that exchanges the particles.) 
Apart from the energy dependence and the factor of two in the denominator, 
this contribution is the same as the one from the corresponding primitive 
orbit for the single-particle density of states 
[Eqs.~(\ref{rough}) and (\ref{isolated})].

As mentioned above, there is also the situation where both particles 
start at the same
point on the orbit and evolve together. Interchanging them
at the end trivially returns them to the same coordinates. This 
pseudo-orbit has action $2S$, but should not be confused with the standard 
dynamical orbit   
where the two particles start at independent points on
the orbit and therefore occur in a 1-parameter family. The fact that we
interchange the particles at the end ensures that the pseudo-orbit 
described here is isolated and does not occur in a family. The
two types of orbits share the same action, but the standard orbit 
has a larger amplitude
due to the different $\hbar$ prefactor and will tend to dominate. This
situation of coexisting contributions with the same action is
analogous to a potential system with a reflection symmetry where
there is a boundary orbit which contributes to both the identity term
in the density of states and also to the reflection term. The only
difference here is that the two types of orbits contribute
with different powers of $\hbar$.

The analysis of the contribution of the type-0 DPPO is similar to
above. We state without proof that its amplitude is simply the same as
the double repetition of the orbit, again divided by
an overall factor of two as discussed in Appendix \ref{determinants}.
This pattern continues for higher repetitions, where for odd multiples
of the action the particles start $T_\gamma/2$ out of phase
while for even multiples they start in phase and interfere with
stronger (in an $\hbar$ sense) contributions from the standard dynamical orbits. 
Apart from the energy dependence and the factor of two in the denominator, 
the sum over 
repetitions is the same as for the single-particle density of states.

Thus, we see that the contribution of the pseudo-periodic orbits 
to the bosonic and fermionic densities of states is precisely the
same as the fluctuating density of states of the one-particle spectrum
except that it is to be evaluated at half the total energy (since the total
energy is partitioned equally between the two particles) and should
also be divided by an overall factor of two. In conclusion,
\be \label{divosc}
\tilde{\rho}_\pm(E) = {1\over 2}
\left(\tilde{\rho}_2(E) \pm {1\over 2}\tilde{\rho}_1 \left({E\over2}\right)
\right),
\ee
consistent with (\ref{symmreddos}).

\subsection{One-particle dynamics in the full phase space}

We now discuss contributions to the resolvent from periodic orbits in the
full phase space where one
particle executes dynamics while the other particle remains
stationary. In particular, suppose that particle $a$ is stationary at
some point in phase space while particle $b$ evolves dynamically on a
periodic orbit. We call this a \textit{heterogeneous} periodic orbit. 
The structure of such orbits is qualitatively different for 
potential systems and billiards.

For analytic potentials, the stationary particle must be at some extremum of
the potential with zero momentum. In this case, the full heterogeneous
orbit is isolated in phase space since a flow in $J=h(z_a)$ does not
map an initial condition $z'$ to any new phase space point
$z$. Therefore, we can use the Gutzwiller trace formula for isolated
orbits.  In billiards, the stationary particle has zero momentum, but
it can be anywhere in the billiard. So rather than
being isolated, the heterogeneous orbits appear as $d$-dimensional
families. This means that we can use the formalism of Ref.~\cite{stephen}
to calculate the amplitude of these orbits. 

The symmetry decomposition for heterogeneous orbits is trivial. Since
the two particles are executing completely different dynamics, the
combination of time evolution and particle exchange, as above, can
never return the particles to their initial conditions. 
This requires an equivalence
of the two motions. Thus, the contribution from heterogeneous orbits is
simply divided evenly between the symmetric and antisymmetric representations
and belongs to the $\tilde{\rho}_2(E)$ term of Eq.~(\ref{divosc}).

\subsubsection{Analytic potentials}\label{potentials}

Suppose particle $b$ traverses a periodic orbit
$\gamma$ with action $S_\gamma$, primitive period $T^0_\gamma$,
stability matrix $\tilde{M}_\gamma$ and topological index
$\sigma_\gamma$. Particle $a$ is assumed to be stationary at a
potential minimum with energy $E_a=0$. At the minimum, the potential
is locally harmonic with $d$ frequencies $\omega_j$. As explained above, 
the full heterogeneous orbit is isolated and so we can
simply use the Gutzwiller trace formula for isolated orbits. The only
required information is the monodromy matrix in the phase space
of particle $a$ since $\det(\tilde{M}_\Gamma-I)=\det(M_a-I)
 \det(\tilde{M}_\gamma-I)$, where $\tilde{M}_\Gamma$ is
the $(4d-2) \times (4d-2)$ stability matrix of the full heterogeneous
orbit and $M_a$ is the $2d \times 2d$ monodromy matrix of particle
$a$. Since the dynamics of particle $a$ are locally harmonic, 
we can use the result
for a $d$-dimensional harmonic oscillator (see Appendix
\ref{appharmonic}), $\sqrt{|{\det}(M_a-I)|} = \prod_{j=1}^d
2\sin(\omega_jT_\gamma(E)/2)$. The phase index of this motion is
simply $d$, one for each transverse harmonic degree of freedom. Thus,
the contribution of one heterogeneous orbit to the resolvent 
is
\begin{equation} \label{scont} \begin{split}
\tilde{g}^{\text{h}}_\Gamma (E)
& = {1\over i\hbar} {T^0_\gamma(E)
\over \sqrt{\left|{\det}\left(\tilde{M}_\gamma-I\right)\right|}
\prod_{j=1}^d 
2\sin\left({\omega_jT_\gamma(E)\over2}\right)} \\
& \qquad \times \exp i {\left({S_\gamma(E)\over\hbar}-
\sigma_{\gamma}{\pi\over2}-d{\pi\over2}\right)}, 
\end{split}
\end{equation}
where we have retained the symbol $\Gamma$ to denote the full 
heterogeneous orbit and $\gamma$ to stress that this is the
contribution from the situation where only one particle is evolving
dynamically. There is also an identical contribution from the
situation where particle $b$ is fixed while particle $a$ evolves
dynamically.  As before, repetitions can be understood to be implicit
in the definitions of the action, period, phase index and stability 
matrix. 

One can also consider extrema other than potential minima, such as
saddles or potential maxima. We can expand the $d$-dimensional
potential around an extremum $x_0$ as 
\be \label{ladedah}
V(x-x_0) = {1\over 2}\left(
\sum_{j=1}^{d_+} \omega_j^2 \xi_j^2 - \sum_{j=d_++1}^d \omega_j^2 \xi_j^2
\right),
\ee
where the $\xi$ measure the deviations of $x$ from $x_0$. In general,
there are $d_+$ stable directions and $d_-=d-d_+$ unstable
directions. Then, the expression (\ref{scont}) is still valid, but the
energy of the dynamically evolving particle is replaced by $E-V(x_0)$,
the phase factor $d\pi/2$ replaced by $d_+\pi/2$ and the
$\sin{(\omega_jT_\gamma/2)}$ replaced by $\sinh{(\omega_jT_\gamma/2)}$
for the unstable directions. Finally, we note that for smooth
potentials, the dynamical
orbits give the leading-order contribution to $\tilde{\rho}_2(E)$ while the
hetero-orbits give corrections of higher order in $\hbar$.  

\subsubsection{Billiard systems}\label{billiardhetero}

As mentioned above, heterogeneous orbits in a $d$-dimensional billiard
occur in $d$-dimensional families and we may therefore use
Eq. (\ref{stephensequation}) with $f=d$ to determine the appropriate
trace formula.  We first consider a two-dimensional billiard ($d=2$), 
although the result is easily generalised. For this
case, the orbits do not appear as three-tori, but rather have the
topology of ${\mathcal{B}}\times {\mathcal{S}}^1$, where ${\mathcal{B}}$ 
denotes the
billiard domain and ${\mathcal{S}}^1$ is the one-torus associated with the
dynamics of the evolving particle $b$ on the periodic
orbit $\gamma$. There are two constants of the motion: $J_1=p_{x_a}$
and $J_2=p_{y_a}$ ($\mathbf{J}=(J_1,J_2)$) and these generate flows 
$\mathbf{\Theta}=(x_a,y_a)$. Clearly,
\be \label{anholobill}
   \det \left({\partial \mathbf{\Theta}\over \partial\mathbf{J}}\right)= 
\det \left( 
{\begin{array}{cc} {\partial x_a\over \partial p_{x_a}} & {\partial x_a\over \partial p_{y_a}} \\ \\
{\partial y_a\over \partial p_{x_a}} & {\partial y_a\over \partial p_{y_a}} 
\end{array}} \right)= {\partial x_a\over \partial p_{x_a}}{\partial y_a\over \partial p_{y_a}}, 
\ee
since the off-diagonal elements vanish due to the fact that the $x$
and $y$ motions are uncoupled.  After particle $b$ has traversed the
primitive orbit $n_\gamma$ times, ${\partial x_a/ \partial p_{x_a}}
={\partial y_a/ \partial p_{y_a}}=-n_\gamma T^0_{\gamma}(E)/m$, where
$T^0_{\gamma}(E)$ is the primitive period of the orbit and $m$ is the
mass of the particle. (The minus
sign indicates that a backwards flow is required to close the orbits
in the full phase space.)  This immediately implies that the phase
index $\delta \equiv 0$.  The stability matrix defined in
(\ref{stephensequation}) in this case is simply the stability matrix
of the motion of particle $b$. The volume for
a family of such orbits is the area of the billiard and combining all
of the factors, the leading-order contribution of a family of
heterogeneous orbits $\Gamma$ to the two-particle density of states is
\be
\tilde{\rho}^{\text{h}}_{\Gamma} (E) = {\alpha {\mathcal{A}}\over 4 \pi^2} 
{\cos\left({S_\gamma(E)\over\hbar}-\sigma_\gamma{\pi\over2}-{\pi\over2}
\right) \over n_\gamma \sqrt{\left|{\det}(\tilde{M}_\gamma-I)\right|}}.
\ee
We obtained this expression in Ref.~\cite{paper1} by doing a direct energy
convolution integral of the first term of Eq.~(\ref{smooth_bill}) with
Eq.~(\ref{rough}). This once again underlines the
equivalence of the two methods. This
naturally extends to the higher order terms of (\ref{smooth_bill})
through a more careful analysis of the surface corrections, but we do
not pursue this analysis here. Also, this result generalises to $d$
dimensions as
\begin{equation} \begin{split}
\tilde{\rho}^{\text{h}}_{\Gamma} (E) & = {1 \over \pi\hbar} 
\left({{\hbar \alpha} \over {4 \pi}}\right)^{d/2}  
{T^0_\gamma(E) \Omega_d \over
\sqrt{\left|{\det}(\tilde{M}_\gamma-I)\right|}
\left({n_\gamma T^0_\gamma(E)}\right)^{d/2}} \\
& \qquad \times \cos\left({S_\gamma(E)\over\hbar}-\sigma_\gamma{\pi\over2}
-d{\pi\over4}\right),
\end{split}
\end{equation}
where $\Omega_d$ is the $d$-dimensional volume of the billiard.  We
stress that this is $O(1/\hbar^{d/2})$ stronger than an
isolated orbit, this factor arising from the fact that this class of
orbits appear in $d$-dimensional families. The contribution from
hetero-orbits is also $O(1/\hbar^{(d-1)/2})$ stronger than the
contribution from dynamical orbits. Thus, for billiards, 
hetero-orbits give the leading-order contribution to $\tilde{\rho}_2(E)$
while dynamical orbits give corrections of higher order
in $\hbar$. 

\section{Several Noninteracting Identical Particles}\label{lotsa}

We now consider the extension to $N$ identical particles. The smooth term
can be written as an $(N-1)$-fold convolution integral of the
single-particle smooth terms and can also be understood as a single
integral in the $N$-particle phase space. At this point, we say no
more about the smooth term and refer the reader to the appendices for
further discussions. For the oscillating term, there are 
again two possibilities.
Either all of the particles are evolving dynamically, or a subset of
them is stationary at various potential extrema (or anywhere in a
billiard).  For the first
situation, the discussion closely parallels the two-particle case. The
only nontrivial quantity to determine is the anholonomy matrix
$\left({\partial\mathbf{\Theta} / \partial \mathbf{J}}\right)_\Gamma$. 
We consider the case $N=3$, but this result readily generalises.

\subsection{Dynamical orbits}

In an obvious extension of the notation, there are three
single-particle phase spaces with coordinates $z_a$, $z_b$ and $z_c$
so that the full three-particle phase space has coordinates
$z=(z_a,z_b,z_c)$ and the total Hamiltonian
$H(z)=h(z_a)+h(z_b)+h(z_c)$. Two other constants of motion which are
in involution with $H$ are $J_a=h(z_a)$ and $J_b=h(z_b)$ and they
generate time translations of particles $a$ and $b$, respectively,
while having no effect on the other particles. 
Flows generated by $H$, $J_a$ and $J_b$ are denoted by $\Phi_t$,
$\Lambda_{\theta_a}$ and $\Psi_{\theta_b}$, respectively. If $\phi$ is
a single-particle flow, then flows in the full phase space are
mapped as follows:
\bea
\Phi_t   (z_a,z_b,z_c) & = & (\phi_tz_a,\phi_tz_b,\phi_tz_c),\nonumber\\
\Lambda_{\theta_a}(z_a,z_b,z_c) & = & (\phi_{\theta_a} z_a,z_b,z_c),\nonumber\\
\Psi_{\theta_b} (z_a,z_b,z_c) & = & (z_a,\phi_{\theta_b} z_b,z_c).
\eea

The periodic orbits of the full phase space (at a given total energy
$E$) can be found from the one-particle periodic orbits by balancing
the energy partition among the three particles (\textit{i.e.} varying
$J_a$ and $J_b$ while holding $H$ fixed) so that all the one-particle
periodic orbits have the same period. (The result is a three-particle
periodic orbit in the full phase space.) Imagine a slight departure
from this equilibrium situation so that $J_a\rightarrow J_a+\Delta
J_a$ while holding $J_b$ and $H$ fixed. Then,
\be
\begin{array}{lcl}
E_a & \rightarrow & E_a + \Delta J_a\\
E_b & \rightarrow & E_b\\
E_c & \rightarrow & E_c - \Delta J_a
\end{array}
\phantom{999}
\begin{array}{lcl}
T_a & \rightarrow & T_a + \Delta T_a\\
T_b & \rightarrow & T_b\\
T_c & \rightarrow & T_c + \Delta T_c
\end{array}
\ee
where $\Delta T_a=T_a'\Delta J_a$ and $\Delta T_c=-T_c'\Delta J_a$,
the primes denoting differentiation with respect to energy. The
initial condition $z'=(z_a',z_b',z_c')$ with these modified energies
(but each particle still on its periodic orbit at that
modified energy) is not on a periodic orbit of the full phase space. 
However, it is on a generalised periodic orbit. That is, the
trajectory can be made to close with additional flows in
$(H,J_a,J_b)$. Imagine a flow in $H$ for the original period $T$. The
orbits of particles $a$ and $c$ will fail to close by the amount by
which their period is longer (or shorter) due to the changed energy: 
$\Phi_T z'= (\phi_{-\Delta T_a}z_a',z_b',\phi_{-\Delta T_c}z_c')$. Additional
flows in $(H,J_a,J_b)$ close the trajectory. First, a flow in $H$ by
the amount $\Delta T_c$ returns particle $c$ to $z_c'$: $\Phi_{\Delta
T_c} \Phi_T z'=(\phi_{-\Delta T_a+\Delta T_c}z_a',\phi_{\Delta
T_c}z_b',z_c')$. The condition for a periodic orbit $\Lambda_{\Delta
{\theta_a}}\Psi_{\Delta {\theta_b}}\Phi_{\Delta T_c} \Phi_T z'=z'$
immediately implies
\bea
\Delta\theta_a & = & (T_a'+T_c')\Delta J_a, \nonumber \\
\Delta\theta_b & = & T_c'\Delta J_a.
\eea

We get a similar result from a deviation in $J_b$ (holding $J_a$ and
$H$ fixed) and conclude that
\be \label{anholo3p}
\left({\partial\mathbf{\Theta}\over\partial \mathbf{J}}\right)
= \left(
\begin{array}{cc}
T_a'+T_c' & T_c'      \\
T_c'      & T_b'+T_c' 
\end{array}
\right).
\ee
The determinant is $T_a'T_b'+T_b'T_c'+T_c'T_a'$ and is invariant under
a permutation of the indices. Note that we could have chosen the two
generators $J_a$ and $J_c$ and followed through the analogous
calculation.  In that calculation, the anholonomy matrix would be
modified by permuting $b$ and $c$ in Eq. (\ref{anholo3p}). Therefore,
the eigenvalues of ${\partial \mathbf{\Theta} / \partial \mathbf{J}}$ are
not invariant. But, since the determinant is invariant, so too is the
number of positive eigenvalues which determines the phase index
$\delta$. Therefore, the final result is invariant. For $N > 3$
particles, this generalises to
\be \label{npartdet}
{\det} \left({\partial\mathbf{\Theta}\over\partial \mathbf{J}}\right) = 
\left(\prod_{p=1}^N T_p'\right) \left(\sum_{p=1}^N{1\over T_p'}\right),
\ee
where $T_p$ is the period of the orbit on which particle $p$
is residing. This can be shown by induction.

The other factors which go into the trace formula are simple to
determine; the discussion is very similar to the two-particle case and
so we refrain from going into great detail. For $N$ particles, flows
in $H$ and $\mathbf{J}=(J_1,...,J_{N-1})$ map out an $N$-dimensional
torus. This means there are $(N-1)$-parameter families of periodic
orbits in the full phase space. The total action is the sum of all the
single-particle actions and similarly for the total phase index $\mu$. The
monodromy matrix is defined holding all of the single-particle
energies constant in such a way that it is block diagonal among the
various single-particle motions. The volume of the periodic orbit
family is the product of the primitive periods. (To see
this, recall that the volume term $T_\Gamma V_\Gamma = \ointop_\Gamma
\d t \d \theta_1 \d \theta_2 \cdots \d \theta_{N-1}$ and that the
primitive volume should only count distinct configurations.) Using
Eq. (\ref{stephensequation}) with $f=(N-1)$, we conclude that the
contribution to the resolvent from one family of dynamical periodic
orbits is
\begin{widetext}
\begin{equation} \label{allevolve} 
 \tilde{g}^{\d}_\Gamma(N,E) =  {1\over i\hbar} 
{1\over(2\pi i\hbar)^{(N-1)/2}}  
\left\{\prod_{p=1}^N {T_p^0(E_p) \exp i \left({S_p(E_p)\over\hbar}-\sigma_p
{\pi\over2}\right) \over
\sqrt{\left|{\det}(\tilde{M}_p-I)\right|}\sqrt{\left|T_p'(E_p)
\right|}}\right\}{\exp i\delta_\Gamma {\pi\over2} \over 
\sqrt{\left|\sum_{p=1}^N{1\over T_p'(E_p)}\right|}}.
\end{equation}
\end{widetext}
In Eq. (\ref{allevolve}), we have used the label $p$ rather than the
more cumbersome $\gamma_p$ to refer to the periodic
orbit on which particle $p$ resides. We will continue to do this for
the remainder of the section. The phase factor $\delta_\Gamma$ is the
number of positive eigenvalues of the $(N-1) \times (N-1)$ matrix
$(\partial \mathbf{\Theta} / \partial \mathbf{J})_\Gamma$. If all of the
particles are on distinct orbits, then there are $N!$
congruent but distinct full phase space orbits, corresponding to the
choice of which particle to assign to which orbit. If
there is more than one particle on the same orbit, then
the number of combinatoric possibilities is accordingly modified. We
take this combinatoric factor to be implicit in the sum over orbits
and do not explicitly account for it here.

\subsection{Hetero-orbits}

The other possibility is that some of the particles are not evolving
dynamically, but rather are stationary in a billiard or at potential
extrema.  Imagine
that $M$ particles are evolving dynamically and $(N-M)$ are fixed at
extrema. Then, these heterogeneous orbits come in $(M-1)$-fold
families. In the special case where the nonevolving particles are
stationary at potential minima,
\be \label{someevolve}
\tilde{g}^{{\h}}_\Gamma(M,N,E) =\tilde{g}^{{\d}}_\Gamma(M,E_e) 
\left\{\prod_{p=M+1}^N{\exp \left(-i{\pi d \over 2}\right) \over
\prod_{j=1}^d 2\sin\left({\omega_{j_p}T\over2}\right)}\right\}.
\ee
The evolving particles share the energy $E_e=E-\sum_{p=M+1}^N
V(x_p)$, where $x_p$ denote the positions of the stationary
particles. We recall that $d$ is the dimension of the one-particle
dynamics and the $\omega_{j_p}$ denote the $d$ local harmonic
frequencies around the minimum at which particle $p$ resides. As in the
two-particle case, if a particle is at a saddle or maximum, we replace
the phase $d\pi/2$ by $d_+\pi/2$, where $d_+$ denotes the number of
stable directions and replace the $\sin$ in the amplitude by $\sinh$
for the unstable directions.  Again, there are distinct but congruent
heterogeneous orbits in which different particles are chosen to be on
different orbits or extrema, but we refrain from an
explicit discussion of the combinatoric possibilities.
 
Next suppose that $(N-M)$ particles are stationary 
in a $d$-dimensional billiard. In addition to the
$(M-1)$ independent generators that exist for the potential system, there are
$(N-M)d$ generators $\mathbf{J_q}=(\mathbf{p}_1,\ldots,\mathbf{p}_{N-M})$.
The associated flows are $\mathbf{\Theta}_\mathbf{q}=(\mathbf{q}_1,\ldots,\mathbf{q}_{N-M})$. (Each $\mathbf{p}$ and $\mathbf{q}$ is $d$-dimensional.) 
Since the
generators associated with the stationary particles also generate new
orbits, the dimensionality of the orbit families is
$f=(M-1)+(N-M)d$. The volume term $T^0_{\Gamma}V^0_{\Gamma}=\oint_\Gamma 
{\d}t {\d}\theta_1 \cdots {\d}\theta_{M-1} {\d}\mathbf{q}_1
\cdots {\d}\mathbf{q}_{N-M}=T^0_1(E_1) \cdots T^0_M(E_M)
\Omega^{(N-M)}_d$. The phase index $\delta_\Gamma$ is the number of
positive eigenvalues of the $f \times f$ matrix $(\partial \mathbf{\Theta} / \partial \mathbf{J})_\Gamma$ which has a block-diagonal
structure; one block is the anholonomy associated with the evolving
particles analogous to Eq.~(\ref{anholo3p}) and one block is the anholonomy
associated with the stationary particles analogous to
Eq.~(\ref{anholobill}). Thus, the contribution to the resolvent from a
family of billiard hetero-orbits is
\be \label{someevolvebill}
\tilde{g}^{{\h}}_\Gamma(M,N,E) = \tilde{g}^{{\d}}_\Gamma(M,E) 
\left\{\prod_{p=M+1}^N{\Omega_d \exp \left(-i {\pi d \over 4}\right) \over
({2 \pi \hbar T \over m})^{d/2}} \right\}.
\ee
In Eqs.~(\ref{someevolve}-\ref{someevolvebill}), $T$ is the global
period (recall that the energies of
all the dynamically evolving particles have been partitioned so that
all of the periodic orbits have a common
period) and $\delta_\Gamma \equiv 0$ if $M=1$.
As in the two-particle case, hetero-orbits are more important in
billiards than in smooth potentials. Their leading-order contribution
to $\tilde{\rho}_N(E)$ is $O(1/\hbar^{(N-M)(d-1)/2})$ stronger for
billiards and $O(\hbar^{(N-M)/2})$ weaker for potentials than the
corresponding contribution from the dynamical orbits. 
 
We now make some final comments. The above expressions apply for any of
the particles executing multiple repetitions of its primitive orbit provided the
energy is partitioned among the dynamically evolving particles so
that all single-particle periodic orbits have a common period. Then, the
various orbit properties which appear in the formulas are understood 
to be those for the repeated orbit. The formulas written above only
account for the contribution of a single family of orbits. The
oscillatory part of the resolvent is a sum over all families:
$\tilde{g}(E)=\sum_\Gamma \tilde{g}_\Gamma(E)$.
We mention that Eqs.~(\ref{allevolve}-\ref{someevolvebill}) can also be obtained
from convolution integrals by doing a stationary phase analysis of the
$N$-particle dynamical term and taking appropriate combinations of
saddle-point and end-point contributions from the various cross-term
integrals. However, the approach outlined above is more illuminating
since it reveals the underlying structure of the periodic orbit families.
The many-particle trace formulas involve only properties of
periodic orbits of the one-particle phase space. Thus, after studying a 
one-particle system, one can immediately work out the details of
the many-particle system. This parallels the situation in quantum
mechanics where the problem of $N$ noninteracting particles in a
potential is a simple extension of the one-particle problem.
  
\section{Symmetry decomposition of the $N$-particle density of states} 
\label{symmN}

If the system consists of $N$ identical particles,
it is invariant under $S_N$, the permutation group
of $N$ identical particles. This group has many different irreps for
$N>2$, but we only consider the one-dimensional bosonic/fermionic 
irreps which are
fully symmetric/antisymmetric under particle exchange. We first introduce the
projection operators \cite{hamermesh}
\be \label{genproj}
\hat{P}_\pm = {1\over N!}\sum_{\tau} (\pm 1)^{n_\tau}
\hat{U}_\tau,  
\ee
where $\pm$ refer to the bosonic/fermionic irreps, respectively. 
The sum is over the group elements $\tau$ of $S_N$ which denote particular
permutations of the particles, $\hat{U}_\tau$ is the representation of
the group element in the Hilbert space (i.e. the quantum operator
which exchanges the particles), $n_\tau$ is the number of 2-particle
exchanges required to obtain $\tau$ and the factor $(\pm 1)^{n_\tau}$
is a group character.  For fermions, the sign of the character depends
on the number of times two particles must be interchanged. As before,
we need to evaluate $g_{\pm}(E)={\Tr}(\hat{P}_{\pm}\hat{G}(E))$ and
therefore ${\Tr}(\hat{U}_\tau\hat{G}(E))$ for each $\tau$. This is
a class function, only depending on the cyclic structure of $\tau$.

Consider a permutation and break it up into cycles
\cite{hamermesh}. For $N$ particles, $\tau$ can be decomposed uniquely
into mutually commuting cycles; in each of these cycles, a subset of
the particles is being permuted. An $n$-cycle is a permutation in
which only $n$ of the particles are being permuted. In particular, a
1-cycle corresponds to an individual particle being left alone, a
2-cycle corresponds to two particles being exchanged with each other
and so on. A general permutation $\tau$ may consist of cycles of
various sizes and also may have several cycles of the same size. In
general, for a given $\tau$, there are $\nu_1$ 1-cycles, $\nu_2$
2-cycles and so on. Then, the cycle structure of a class of
permutations can be given as a set of integers 
$\left(\nu_1, \nu_2, \ldots, \nu_N \right)$. This set
$\bm{\nu}$ labels the conjugacy classes. Two permutations 
with the same $\bm{\nu}$ belong to the same
class and thus have the same value of ${\Tr}
(\hat{U}_\tau\hat{G})$. The analysis of the previous section can be
understood as being the special case of the identity element.
To decompose the full density of states, one needs to determine both
the smooth and oscillating contributions to 
${\Tr} (\hat{U}_\tau\hat{G})$. The smooth contribution is discussed
in Appendix \ref{TFNpsym}. In this section, we examine the oscillating
contribution.

\subsection{Dynamical cycles}\label{dyncycles}

Consider first the case for which all particles are evolving dynamically. A
group element $\tau$ consists of $m_\tau$ cycles, a given cycle $k$
consisting of interchanging $n_k$ particles. As in the two-particle
case, particle interchange does not commute with all of
the single-particle energies and so we do not expect periodic orbit
families of dimension $(N-1)$. However, for each cycle, there is a
generator $J_k$ which is the sum of the single-particle Hamiltonians
of the particles involved in that cycle and is preserved under the
action of the group element $\tau$. These generators commute with each
other and with the total Hamiltonian $H$. However, this is not an
independent set since $\sum_k J_k=H$. There are $(m_\tau-1)$
independent commuting generators other than the full Hamiltonian and
so we expect periodic orbit families of this dimensionality
contributing to ${\Tr}(\hat{U}_\tau\hat{G})$.

\begin{figure}
\scalebox{0.33}{\includegraphics{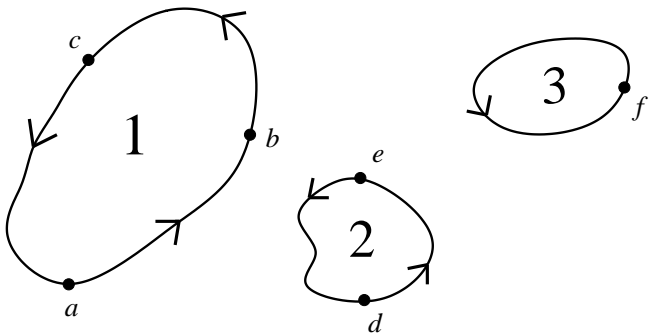}}
\caption{\label{cycle123} A specific permutation of $6$ particles 
is decomposed into
three dynamical cycles. Each of the particles belonging to a particular cycle
are on a periodic orbit of the one-particle phase space with
$T_1(E_a=E_b=E_c \equiv E_1)/3 = T_2(E_d=E_e \equiv E_2)/2= T_3(E_f
\equiv E_3) \equiv T$.}
\end{figure}

We seek structures in the full phase space which are invariant under
the combined operations of time evolution (for time $T$) generated by
$H$ and particle exchange as specified by $\tau$. Clearly, this
is only possible if all particles of a given cycle $k$ are on
the \emph{same} periodic orbit, $\gamma_k$. They must all have the same
energy, which we shall call $E_k$ and then $J_k=n_kE_k$. For example,
imagine that particles $a$, $b$ and $c$ constitute a 3-cycle. Starting
with particle $a$ at some arbitrary point on a periodic orbit
$\gamma$ of the one-particle phase space, particle $b$ an amount $T_\gamma/3$ ahead of it and particle
$c$ an amount $T_\gamma/3$ behind. Then, after a time $T=T_\gamma/3$,
$a \rightarrow b$, $b \rightarrow c$ and $ c\rightarrow a$. However,
the group element $\tau=(acb)$ maps $a \rightarrow c$, $c \rightarrow b$ and 
$b\rightarrow a$ simply undoes this change and the original
configuration is restored. Such a cycle is shown at the left of
Fig.~\ref{cycle123}. We then imagine that for every cycle comprising
$\tau$, there is a train of particles with identical energies
traversing a periodic orbit of the one-particle phase space. 
Each particle completes $(1/n_k)$th of the full motion on the periodic orbit.

The logic then is as follows. We assign each cycle a periodic orbit
$\gamma_k$. (We will henceforth label the orbit properties using the
subscript $k$ rather than the more cumbersome $\gamma_k$.) We
partition the energy (\textit{i.e.} the values of $J_k$) so that the
values of the periods $T_k/n_k$ are all the same; this quantity we
denote by $T$. After time
$T$ and permutation $\tau$, the resulting structure is guaranteed to
be globally periodic in the full phase space. Such an orbit comes in
an $(m_\tau-1)$ degenerate family which can be understood as
follows. For each cycle, it is enough to specify the initial condition
of one particle after which we know the initial conditions of all the
other particles. We choose the initial condition of the first particle
arbitrarily for the first cycle. The first particle of the other
$(m_\tau-1)$ cycles can then begin anywhere on their respective orbits
(this constituting the dimensionality of the family). We also
understand this from the fact that starting at the arbitrary initial
condition, flows generated by any of the $(m_\tau-1)$ generators $J_k$
map out a surface of this dimensionality. Together with a flow in
$H$, the periodic orbit surface is a torus of dimension $m_\tau$.

\begin{figure*}
\hspace*{-1.9in} 
\scalebox{0.35}{\includegraphics{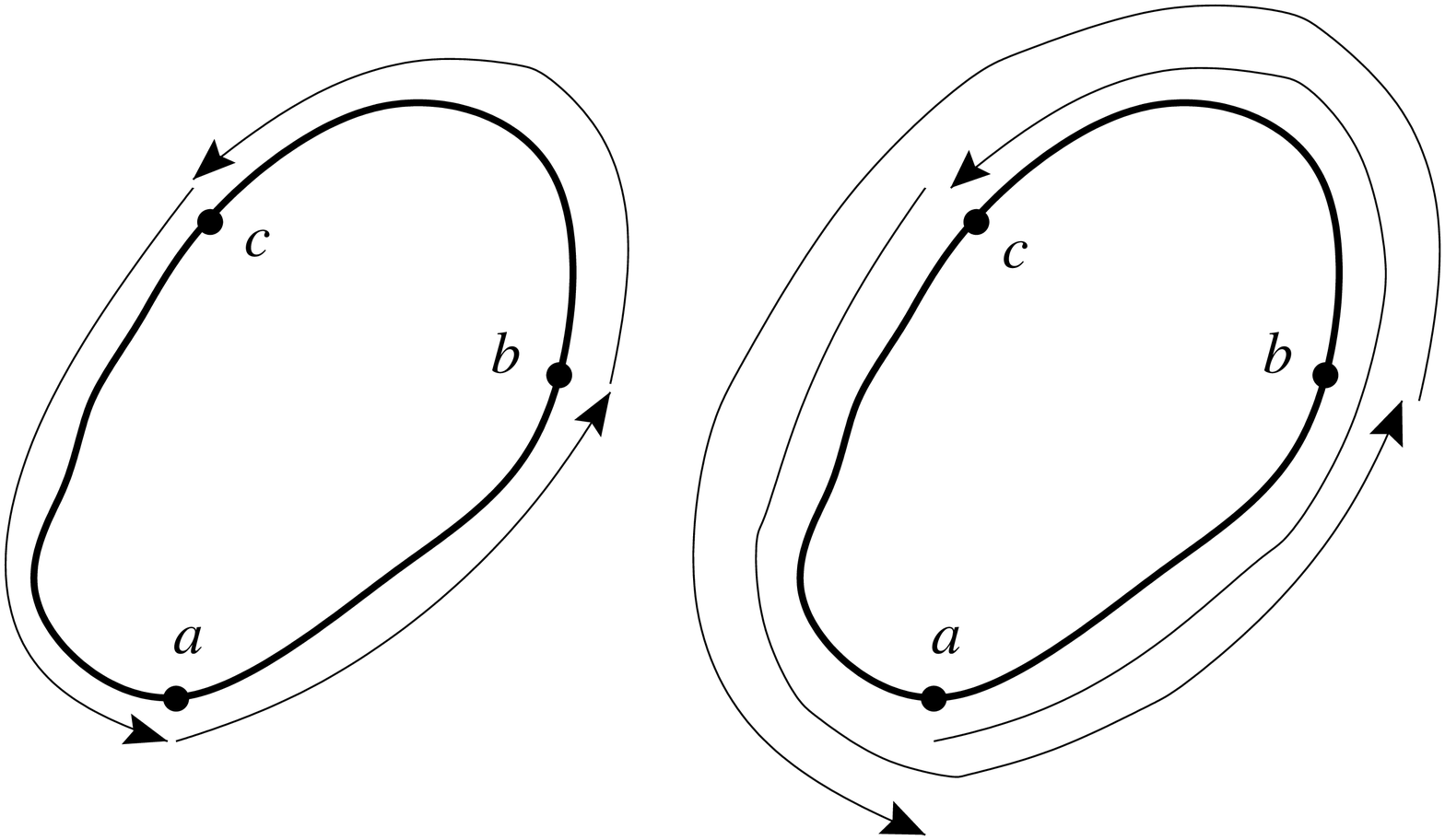}}
\vspace*{-5.25cm}
\begin{center}
\hspace*{3.75in}
\scalebox{0.35}{\includegraphics{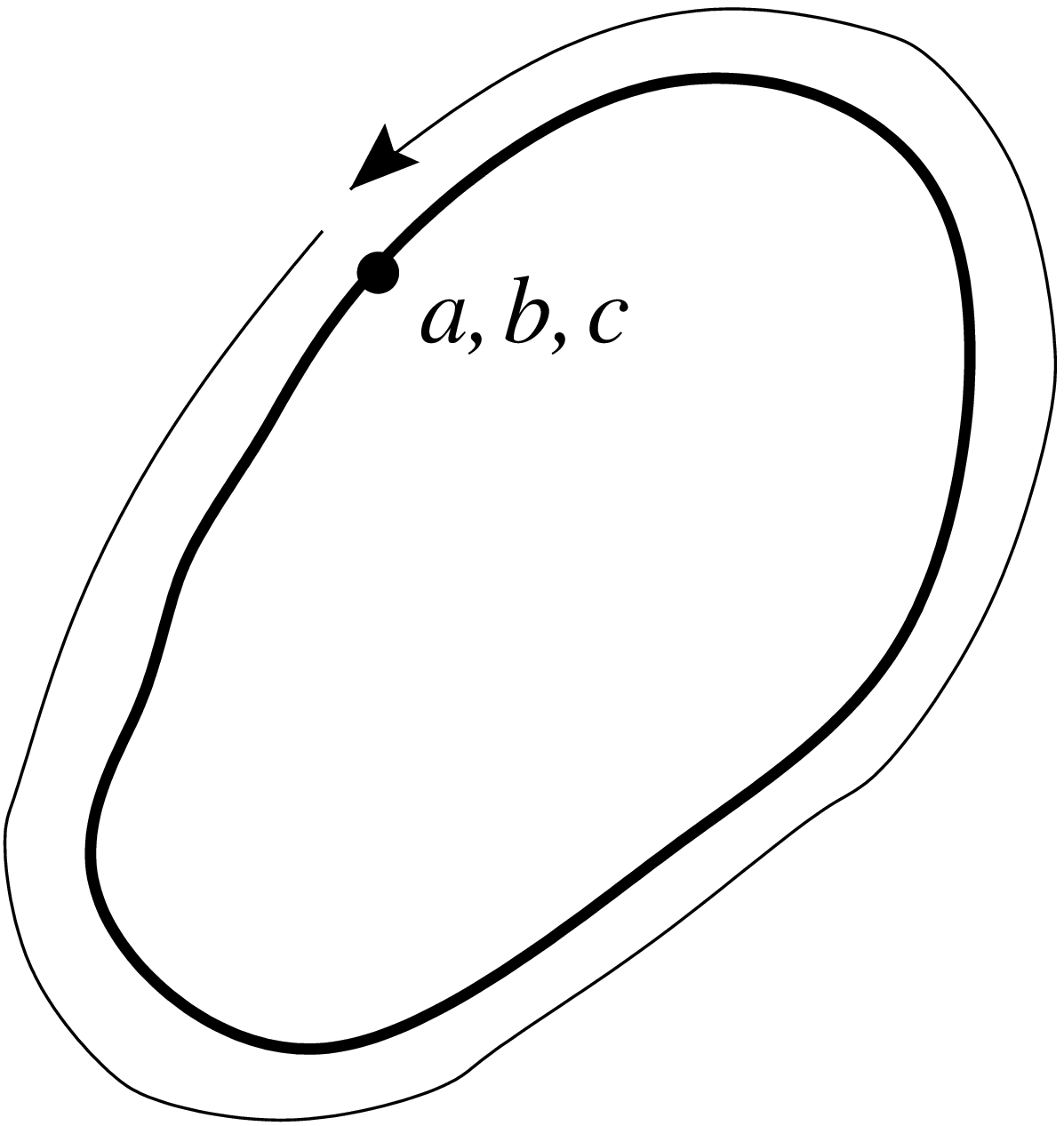}}
\end{center}
\caption{\label{uglyfig} (Left) The same type-1 dynamical 3-cycle of 
Fig.~\ref{cycle123}. 
(Middle) A type-2 dynamical 3-cycle. 
The same periodic orbit, but each particle executes two-thirds of 
the complete motion and the net action is $2S$. (Right) A type-0 dynamical 
3-cycle. Each particle executes one complete motion and the net action is $3S$.}
\end{figure*}

For the symmetry decomposition (involving the dynamical orbits) of
a two-particle system, it was noted that there were contributions from
higher multiples. For instance, one could start both particles at the
same point on a single-particle orbit, let them evolve for a full
period and then interchange them. There is an analogous
structure in the $N$-particle case. We can allow the
particles to execute a fraction $l_k/n_k$ of an orbit as depicted in
Fig.~\ref{uglyfig}. As before, the additional factor $l_k$ can be
absorbed into the definitions of the various classical parameters. 

The contribution of an $m_\tau$-torus of orbits can now be inferred from our
previous work. The only detail is in the determination of
$\left({\partial\mathbf{\Theta} / \partial \mathbf{J}}\right)$. It is as in
(\ref{npartdet}), but with the understanding that the sum/product over
orbits should be replaced by a sum/product over cycles. These
become equivalent in the identity contribution which was considered
there. Also, since the anholonomy term measures deviations away from
global periodicity arising from a change in the energy partition (now
among the cycles), $T_p'$ should be replaced by $T_k'/n_k^2$. A factor
of $1/n_k$ comes from the fact that the energy of the cycle must be
divided evenly among the $n_k$ particles belonging to that cycle. A
second factor of $1/n_k$ comes from the fact that the orbit has time
$T_k/n_k$ for the anholonomy to evolve. (Note that if this orbit is a
multiple repeat, then it is understood that $T_k'=l_kT_k^{0'}$, where
$T_k^0$ is the primitive period.) The entire contribution should also
be divided by $\prod_k n_k$ arising from the monodromy matrix as
discussed in Appendix~\ref{determinants}. This last fact is the
generalisation of the factor of $1/2$ appearing as a prefactor in the
second term of Eq.~(\ref{divosc}) for the two-particle case.
Therefore, the contribution from a family of dynamical cycles to ${\Tr}
(\hat{U}_\tau\hat{G})$ can be written as
\begin{widetext}
\begin{equation} \label{fullsymallevolve} 
 \tilde{g}^{\d}_\tau(m_{\tau},E) =  {1\over i\hbar} {1\over(2\pi
i\hbar)^{(m_\tau-1)/2}}  
\left\{\prod_{k=1}^{m_\tau} {T_k^0(E_k) \exp i \left({S_k(E_k)\over\hbar}-
\sigma_k{\pi\over2}\right) \over \sqrt{\left|{\det}
(\tilde{M}_k-I)\right|}\sqrt{\left|T_k'(E_k)\right|}}\right\}
{\exp i \delta {\pi\over2} \over\sqrt{\left|\sum_{k=1}^{m_\tau}{n_k^2\over
T_k'(E_k)}\right|}},
\end{equation}
\end{widetext}
where $\tilde{M}_k$ is the stability matrix for a full cycle $k$
(cf. Appendix \ref{determinants}). Notice that the contribution of the 
group element for which \emph{all} of the particles belong 
to the same cycle   
is proportional to $\tilde{\rho}_1(E/N)$ \footnote{The
division by $N$ of the energy argument simply states that the total energy
must be \emph{evenly} divided among all of the particles. The set of orbits
corresponding to this cycle is clearly the same as the set of orbits
of the one-particle dynamics (almost by definition) and the
amplitudes and actions are the same as the one-particle case since
the $N$ particles \emph{collectively} execute one complete motion (or 
a multiple repetition) of the periodic orbit. 
This is observed for $N=3$ in section \ref{symdecompforcard}.}.

\subsection{Hetero-cycles}

It is also possible that ${\Tr}(\hat{U}_\tau\hat{G})$ has a
contribution from cycles where some particles are fixed (either at
extrema of the potential or anywhere in a billiard) while others are evolving dynamically. Let
$s$ denote the number of cycles that are stationary and $e$ the
number of cycles that are evolving dynamically. Then,
$s+e=m_\tau$. To have such a contribution to the oscillating
component, group elements must consist of two or more cycles since 
those that consist of only one $n_k$-cycle will
either contribute to Eq.~(\ref{fullsymallevolve}) if they are
dynamical cycles or to the smooth part (cf. Appendix \ref{TFNpsym}) if
they are stationary cycles. In addition, 
we require at least one cycle to involve particles which
are evolving dynamically ($e \ge 1$) \emph{and} at least one
cycle to involve particles which are stationary ($s \ge 1$). Thus,
hetero-cycles are cycles for which $1 \le e < N$ and $1
\le s < N$.

For potentials, the dimension of a family of
orbits is then $(e-1)$ since only the generators associated with
dynamical cycles generate new orbits. The stationary cycles simply
contribute their monodromy matrices and phase indices and otherwise
play no essential role. Eq.~(\ref{fullsymallevolve}) holds for the particles
which are evolving dynamically, but $m_\tau$ is replaced by $e$
and the energy associated with the $e$ dynamical cycles, $E_e$, is the total
energy minus the sum of the potential energies of the stationary
particles. For a potential minimum, the contribution of one 
family of hetero-cycles to ${\Tr}(\hat{U}_\tau\hat{G})$ is
\be \label{thelast}
\tilde{g}^{\h}_\tau(e,m_\tau, E)=\tilde{g}^{\d}_\tau
(e,E_e)  
\left\{\prod_{k=e+1}^{m_\tau} {\exp \left(-i {\pi d \over 2}\right) \over
\prod_{j=1}^d 2\sin\left({\omega_{j_k}T_k\over2}\right)}\right\},
\ee
where the $\omega_{j_k}$ denote the local frequencies around the
potential minimum at which the particles of cycle $k$ reside. If this
cycle of particles is actually at a saddle or a maximum, the final
factor is modified as in the discussion below Eq.~(\ref{ladedah}).

For billiards, the previous relation holds for the dynamical cycles, but
the product over stationary cycles is modified. As explained below, 
the dimension of the orbit families is $[(e-1)+sd]$ 
since the generators associated with stationary cycles also
generate new orbits. If there are $s$ stationary 1-cycles, these
generators and their flows are $\mathbf{J}=(\mathbf{p}_1,\ldots,\mathbf{p}_s)$
and $\mathbf{\Theta}=(\mathbf{q}_1,\ldots,\mathbf{q}_s)$, respectively. (There are
$sd$ components since each $\mathbf{p}_i$ and $\mathbf{q}_i$ is 
$d$-dimensional.) In fact,
this is true regardless of the number of particles belonging to the 
stationary cycle. At
first, this may seem incorrect since longer cycles will introduce additional
generators because they involve more particles. However, this
larger set of generators is not an independent set. To see this,
recall that the particles involved in a stationary cycle can be
anywhere in the billiard. If the cycle is not a 1-cycle, but rather an
$n_k$-cycle, the combined operations of time evolution and particle
exchange will not restore the initial configuration unless all the
particles involved in that cycle possess the same phase
space coordinates. More formally, a stationary
$n_k$-cycle possesses a set of generators, $\mathbf{J}=(\mathbf{p}_1,\ldots,\mathbf{p}_{n_k}$) and associated flows $\mathbf{\Theta}=(\mathbf{q}_1,\ldots,\mathbf{q}_{n_k}$), where each $\mathbf{p}_i$ and $\mathbf{q}_i$ has $d$ components. However,
after the specification of a single
($\mathbf{p}_i$,$\mathbf{q}_i$) pair, all others are uniquely determined: 
$\mathbf{p}_1=\mathbf{p}_2=\cdots=\mathbf{p}_{n_k}$ \emph{and} $\mathbf{q}_1=\mathbf{q}_2=\cdots=\mathbf{q}_{n_k}$. Thus, one independent set of generators 
is $\mathbf{J}=\mathbf{p}_i/{n_k}$, where $\mathbf{p}_i$ is the momentum of the $i$th
particle of the stationary cycle. 
The factor of $1/{n_k}$ comes from the fact
that the momentum of the cycle must be equally partitioned among the
$n_k$ particles belonging to that cycle. We note here that it is not
necessary for stationary particles of distinct cycles
to have the same phase space coordinates. 
Thus, for a $d$-dimensional billiard, the
contribution to ${\Tr}(\hat{U}_\tau\hat{G})$ from a family of 
hetero-cycles is   
\be \label{thelastbill}
\tilde{g}^{\h}_\tau(e,m_\tau, E)=\tilde{g}^{\d}_\tau(e, E) 
\left\{\prod_{k=e+1}^{m_\tau} {\Omega_d \exp\left(-i{\pi d \over 4}\right) \over
\left({2\pi\hbar n_kT_k\over m}\right)^{d/2}}\right\}.
\ee
Eqs.~(\ref{thelast}-\ref{thelastbill}) are the most general formulas 
of the paper. We allow for any amount
of particle permutation and any number of particles can be evolving
while the rest are stationary. Each cycle can involve an arbitrary
repetition of the primitive motion. As before, if $e=1$, 
then $\delta \equiv 0$. 
Hetero-cycles in billiards are $O(1/\hbar^{sd/2})$ stronger than in smooth
potentials. For potentials [billiards], hetero-cycles are
$O(\hbar^{s/2})$ weaker [$O(1/\hbar^{(d-1)s/2})$ stronger]
than dynamical cycles. Thus, the most significant
structures (in an $\hbar$ sense) are dynamical cycles for smooth potentials and
hetero-cycles for billiards. 

\section{Numerical Example: The Three-Particle Cardioid Billiard}\label{num3pp}

To illustrate the use of the trace formulas derived above, we study a system 
of three noninteracting identical particles in a
two-dimensional cardioid billiard. In a billiard, 
classical orbits possess simple scaling
properties. For instance, the action and period of an orbit $\gamma$
are 
\bea \label{Tbill}
S_{\gamma}(\varepsilon) & = & \sqrt{2 m \varepsilon} L_{\gamma}, \nonumber \\
T_{\gamma}(\varepsilon) = S_{\gamma}'(\varepsilon)
& = & {\sqrt{2m} L_\gamma \over {2 \sqrt{\varepsilon}}}
= {\hbar\sqrt{\alpha}\over 2\sqrt{\varepsilon}}L_\gamma.
\eea
For this reason, it is natural and convenient to analyse the length
spectrum of the various trace formulas. Thus, for our analysis,
we shall compare Fourier transforms
of quantum spectra with their semiclassical approximations 
in the reciprocal space of orbit lengths, $L$.  
In reciprocal $L$-space, we expect signals at the lengths of the full
periodic orbits of the three-particle system. In the subsequent
analysis, peaks in the various length spectra are identified with
particular periodic orbits of the full classical phase space. 
We first consider the total (three-particle) density of states for the
cardioid system and then study its decomposition among the irreps
of $S_3$. 

\subsection{Total density of states}
\subsubsection{Quantum mechanics}

The analogue of Eqs.~(\ref{1pand2pqndos}) and (\ref{2ppconv}) for
the quantum three-particle density of states is 
\begin{eqnarray}
   \rho_3(E) & = &  \sum_{i,j,k}
   \delta(E-(\epsilon_{i}+\epsilon_{j} +\epsilon_{k})) \nonumber \\
    & = &  \rho_1(E) * \rho_1(E) * \rho_1(E).
\end{eqnarray}
In fact, this relation applies even if
the particles are not identical where the full density is still the
convolution of the three distinct single-particle densities. 
We construct the three-particle spectrum by adding 
the energies of the one-particle spectrum. (The billiard has a
reflection symmetry which implies that all the single-particle states
are either even or odd; this symmetry should not be confused with the
symmetry due to particle exchange.) In the subsequent analysis, we 
work exclusively with the odd-parity one-particle spectrum.  
We include the first 500 
single-particle energies which allows us to construct the first 19,317,062
energy levels representing all three-particle energies
less than $2.8148 \times 10^3$. (The spectrum is truncated at $E_{\text{max}}
=2 \epsilon_1 + \epsilon_{500}$ to ensure there are no missing levels.)
It is possible to improve the resolution in $L$-space by truncating 
the spectrum at a higher energy.  But, this would require a 
precise spectrum since there is a rapid 
increase in the number of three-particle levels with energy and 
errors accumulate. 
   
\subsubsection{Weyl term}

Using the identity contribution from
Appendix \ref{TFNpsym}, the smooth three-particle density of states
is just the 2-fold convolution integral of the
smooth single-particle density of states:
\begin{eqnarray}\label{3psmooth}
\bar{\rho}_3(E)=\bar{\rho}_1(E) * \bar{\rho}_1(E) * \bar{\rho}_1(E).
\end{eqnarray}
For a two-dimensional billiard, we use Eq.~(\ref{smooth_bill}) for
$\bar{\rho}_1(E)$. After performing the necessary integrations
(ignoring terms $O(1/\hbar^3)$), the 
three-particle smooth term is found to be
\begin{equation}\label{3psmthfin} \begin{split}
\bar{\rho}_3(E) & =
{\alpha^3{\mathcal{A}}^3 \over 128 \pi^3}E^2 -{\alpha^{5/2}{\mathcal{A}}^2
{\mathcal{L}} \over 32 \pi^3}E^{3/2} \\
& \qquad +{3  \over 2}\alpha^2  \left({{\mathcal{A}}{\mathcal{L}}^2 \over 128 \pi^2}+{{\mathcal{A}}^2 {\mathcal{K}} \over 16 \pi^2}\right)E.
\end{split}
\end{equation}
For the odd-parity
single-particle  spectrum of the cardioid,
${\mathcal{A}}=3\pi/4$, ${\mathcal{L}}=6$, and ${\mathcal{K}}=3/16$.
Some of the contributions of the higher-order terms of
$\bar{\rho}_3(E)$ can be calculated, but 
it is formally meaningless to include them 
since there are corrections of the
same relative order in $\hbar$ that are not known. 
The terms that are $O(\sqrt{\alpha^3E})$ and  
$O(\alpha E^0)$ can be computed numerically.
 
\subsubsection{Hetero-orbits}\label{hetero3ppbill}

For three particles in a two-dimensional billiard, there are two types
of heterogeneous orbits. The first type occurs when one particle
is on a periodic orbit while the other
two particles are stationary. These orbits come in 4-parameter families.
The trace formula is obtained by using Eq.~(\ref{someevolvebill}) with
$M=1,N=3$. For the situation where particles $a$ and $b$ are stationary
and particle $c$ evolves on the orbit $\gamma$, the leading-order
contribution to $\tilde{\rho}_3(E)$ is 
\begin{equation} \label{H1} \begin{split}
 \tilde{\rho}^{\text{h1}}_3(E) & = 
 {{\alpha^{3/2}{\mathcal{A}}^2 E^{1/2}} \over {8 \pi^3}} \sum_\gamma
     {{(L^0_\gamma/L^2_\gamma)} \over {\sqrt{\left|{\det}
(\tilde{M}_\gamma-I)\right|}}} \\ 
& \qquad \times \cos \left(\sqrt{\alpha E}L_\gamma -
     {\sigma_\gamma}{\pi \over 2}-\pi \right).
\end{split}
\end{equation}
The second type of hetero-orbit arises from the situation
where only one particle is stationary while the other two evolve
on periodic orbits. For instance, particle $c$ is
stationary while particle $a$ evolves on $\gamma_a$ and particle $b$
evolves on $\gamma_b$. Using formula (\ref{someevolvebill}) with
$M=2,N=3$, we conclude the leading-order contribution to
$\tilde{\rho}_3(E)$ from these hetero-orbits is 
\begin{equation} \label{H2} \begin{split}
& \tilde{\rho}^{\text{h2}}_3(E)   = 
{{\alpha^{5/4}{\mathcal{A}} E^{1/4}} \over 
{{(2 \pi)}^{5/2}}} \sum_{{\gamma_a},{\gamma_b}}
\left({\prod_{p=a,b}}{{L^0_{\gamma_p}} \over 
{{\sqrt{\left|{\det}(\tilde{M}_{\gamma_p}-I)\right|}}}}\right) \\
& \quad \times 
{\left(L^2_{\gamma_a}+L^2_{\gamma_b}\right)}^{-3/4}
\cos \left(\sqrt{\alpha
     E} L_{\Gamma} - {\sigma_{\Gamma}}{\pi \over 2}-{3 \pi \over 4}
     \right),
\end{split}
\end{equation}
where $L_{\Gamma}=\sqrt{L^2_{\gamma_a}+L^2_{\gamma_b}}$ and
$\sigma_{\Gamma}=({\sigma_{\gamma_a}+\sigma_{\gamma_b}})$.

For the total density of states, both formulas are multiplied by a 
factor of $3$ since
there are three identical contributions
depending on the choice of which particle is evolving and which is
stationary. Higher order contributions can be obtained using the
convolution formalism and the results are given in Appendix
\ref{conv3pp}.

\subsubsection{Dynamical orbits}

To use formula (\ref{allevolve}), we must first determine the energies that 
satisfy the following conditions:
\begin{eqnarray} \label{dyn3ppcon}
& T_{\gamma_a}(E_a)=T_{\gamma_b}(E_b)  =  T_{\gamma_c}(E_c), \nonumber \\
& E_a + E_b  +  E_c = E. 
\end{eqnarray}
This leads to a simple linear system which can be solved to give 
\begin{eqnarray} \label{3ppen}
 E_i=\left({{L^2_{\gamma_i}}\over 
{L^2_{\gamma_a}+L^2_{\gamma_b}+ L^2_{\gamma_c}}}\right) E 
\end{eqnarray}
for $i=a,b,c$.
We can now proceed to compute each of the quantities involved in
formula (\ref{allevolve}). The anholonomy term (cf. Eq.~(\ref{anholo3p})) 
\begin{eqnarray} \label{anholo3pp}
{\det} \left({\partial\mathbf{\Theta}\over\partial \mathbf{J}}\right) & = &
T_{\gamma_a}'T_{\gamma_b}'+T_{\gamma_b}'T_{\gamma_c}'+T_{\gamma_c}'T_{\gamma_a}' \nonumber \\
 & = & {{\hbar^2 \alpha} \over 16}{{L^8_{\Gamma}} \over
{L^2_{\gamma_a}L^2_{\gamma_b}L^2_{\gamma_c}E^3}}.
\end{eqnarray}
In addition, ${\Tr} \left({\partial\mathbf{\Theta}\over\partial \mathbf{J}}\right) < 0$ and 
this implies the phase factor $\delta_\Gamma \equiv 0$. Then, the
three-particle dynamical term can be written as
\begin{equation}\label{dynterm3pp}\begin{split}
\tilde{\rho}^{\d}_3(E) & = {\alpha \over {\left(2 \pi
\right)^2}} \sum_{{\gamma_a},{\gamma_b},{\gamma_c}} {1 \over
{L_{\Gamma}}} \left({\prod_{p=a,b,c}}{{L^0_{\gamma_p}} \over 
{{\sqrt{\left|{\det}(\tilde{M}_{\gamma_p}-I)\right|}}}}\right) \\  
& \qquad \times \cos \left( \sqrt{\alpha E} L_{\Gamma}-\sigma_{\Gamma}{\pi
\over 2}-{\pi \over 2}\right),
\end{split}
\end{equation}
where $L_{\Gamma}=\sqrt{L^2_{\gamma_a}+L^2_{\gamma_b}+L^2_{\gamma_c}}$ and
$\sigma_{\Gamma}=({\sigma_{\gamma_a}+\sigma_{\gamma_b}+\sigma_{\gamma_c}})$.

\subsubsection{Numerics}
We first mention that for billiards, it is common to express the
density of states in terms of the wavenumber $k$, where $\varepsilon =
k^2/\alpha$ so that $\rho(k)=2k\rho(\varepsilon)/\alpha$. This is
convenient since $k$ is conjugate to the
periodic orbit lengths $L$. Therefore, our numerical results will be
quoted as functions of $k$ with the understanding 
that these functions have been converted to the $k$ domain from the
energy domain using
the Jacobian relation above. This will always be the case when
the argument is $k$. As well, for all numerical comparisons, $\alpha$
(and $\hbar$) are set to unity. 

\begin{figure}
\scalebox{0.533}{\includegraphics{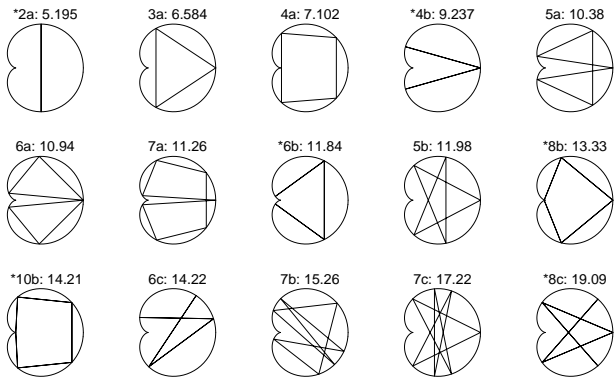}}
\caption{\label{Bruusfig} Some of the shorter periodic orbits of the cardioid
in the full domain. The label of each orbit includes the number of
reflections and also a letter index to further distinguish it. The asterisk designates a self-dual orbit \cite{Bruus}. The two
orbits *8b and *10b reflect specularly near the cusp, contrary to
appearances while the orbit 4a misses the cusp. From Ref.~\cite{Bruus}.}
\end{figure}

\begin{figure}
\hspace*{-0.225in} 
\scalebox{0.5}{\includegraphics{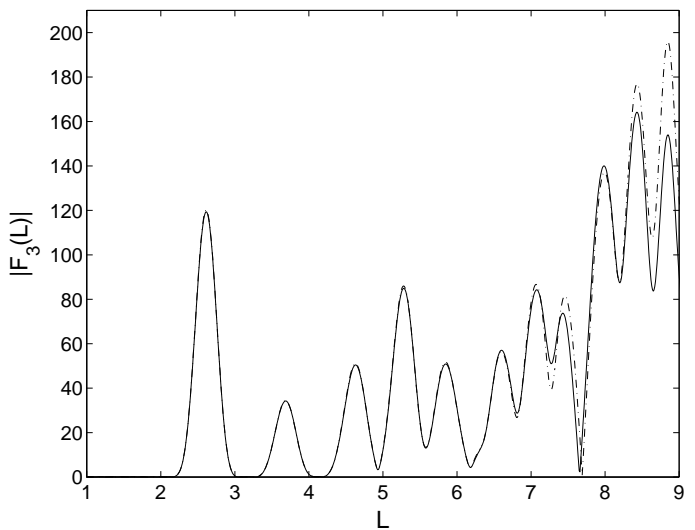}}
\caption{\label{full3pp} Fourier transform of the oscillatory part of the
three-particle density of states for $L < 9$. The solid line is the 
transform of the quantum three-particle spectrum (\ref{qn3pp}) and the 
dashed-dotted line is the transform of the combined semiclassical
three-particle trace formulas (\ref{osc3pp}). Each structure is due to one or
several periodic orbits of the full phase space.} 
\end{figure}

We compare the Fourier transform of
the oscillatory part of the density of states  
\begin{eqnarray}\label{osc3pp}
\tilde{F}_{3}^{\text{sc}}(L) & = & 
{\mathcal{F}} \{ \tilde{\rho}_3(k) \} \nonumber \\  
    & = & {\mathcal{F}} \{ 3
    \tilde{\rho}^{\text{h1}}_3(k) +
    3 \tilde{\rho}^{\text{h2}}_3(k) + \tilde{\rho}^{\d}_3(k) \}  
\end{eqnarray}
with its quantum mechanical analogue which we define to be
\begin{eqnarray}\label{qn3pp}
   \tilde{F}_{3}^{\text{qm}}(L)={\mathcal{F}}\{\rho_3(k) - \bar{\rho}_3(k)\}.  
\end{eqnarray}
In Eq.~(\ref{qn3pp}), the first term is the quantum three-particle
density of states, $\rho_3(k) = \sum_{I}
\delta \left(k - k_I \right)$, where the superindex $I$ denotes a
triplet of integers ($i,j,k$). The subtracted term is the smooth
density of states as determined from Eq.~(\ref{3psmthfin}). 
The oscillatory part has contributions from hetero-orbits
(\ref{H1}-\ref{H2}) and dynamical orbits (\ref{dynterm3pp}). In all
formulas, $\gamma_p$ are periodic orbits in the fundamental
domain (i.e. the half-cardioid) and $L^0_{\gamma_p}$ are their primitive
lengths. Orbit properties are discussed in Refs.~\cite{Bruus,Backer} 
and some of
the shorter geometrical orbits are shown in Fig.~\ref{Bruusfig}. 
The stability matrices of the Gutzwiller amplitudes are
computed using the standard procedure for the stability of free-flight
billiards (see, for example, Ref.~\cite{dasbuch}). 
We define the Fourier transform
\be\label{Fourier}
{\mathcal{F}}\{\rho(k)\} = \int_{-\infty}^{\infty} {\text{w}}(k) 
\exp(ikL) \rho(k) \d k 
\ee
as a function of the conjugate variable $L$. Here, ${\text{w}}(k)$
is the three-term Blackman-Harris window function \cite{Harris}
\be\label{BHW1}
{\text{w}}(k) = \left\{\begin{array}{lrl} \sum_{j=0}^{2} a_j \cos \left(
{2 \pi j} {{k-k_0} \over {k_f-k_0}} \right) && k_0 < k < k_f \\  
0 && \textrm{otherwise}
\end{array} \right.
\ee
with $(a_0,a_1,a_2) = (0.42323,-0.49755,0.07922)$. We choose $k_0$ and
$k_f$ so that the window function goes smoothly to zero at the first
and last eigenvalues of the three-particle spectrum.  Numerical
integration of (\ref{osc3pp}) and (\ref{qn3pp}) using this
integral operator is displayed in Fig.~\ref{full3pp}. In the
semiclassical transform, a total of 212 periodic orbits (including
multiple repetitions) were used. 

\begin{figure}
\hspace*{-0.225in} 
\scalebox{0.45}[0.533]{\includegraphics{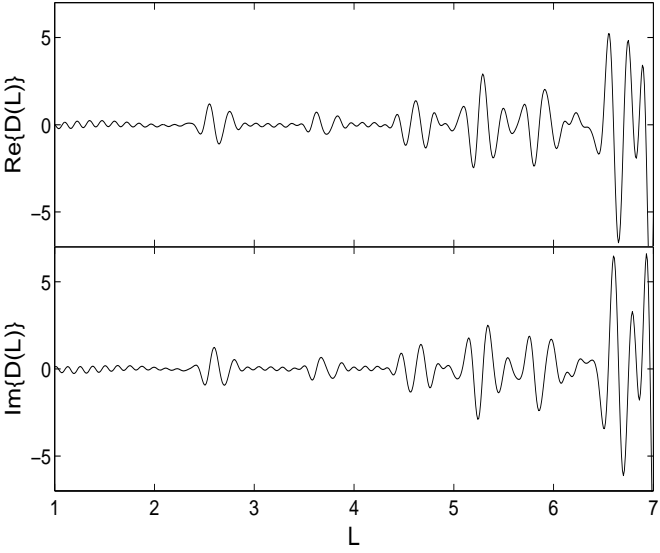}}
\caption{\label{diff3pp} Fourier transform of the difference between the 
quantum and semiclassical density of states for $L < 7$. The upper
and lower windows show the real and imaginary parts, respectively.} 
\end{figure}

We observe good agreement between quantum
and semiclassical results for $L<7$.  In fact, it is difficult to
distinguish between the two curves. For this reason, we plot the
difference between them in Fig.~\ref{diff3pp}. Clearly, the
errors are small with respect to individual peak heights. Furthermore,
the errors are largely due to hetero-orbit contributions. This can be 
understood by considering the first three structures in $L$-space. The first
structure ($L \approx 2.60$) is due to a type-1 hetero-orbit where two
particles are stationary and one evolves on
$\gamma={1\over2}(^*$2a). The second structure ($L \approx 3.67$) 
is from a type-2 hetero-orbit
where one particle is stationary and two particles evolve
independently on the same orbit $\gamma={1\over2}(^*$2a). 
The third structure arises from the interference between a
type-1 hetero-orbit ($L \approx 4.62$) where one of the three particles is on
$\gamma={1\over2}(^*$4b) and a dynamical orbit ($L \approx 4.50$) 
where all three
particles evolve independently on $\gamma={1\over2}(^*$2a). We see
that the first and third structures have similar errors and thus 
conclude that the error introduced from the dynamical term is much
smaller than that from the heterogeneous terms. All other $L$-space
structures arise from the interference of many orbits and 
can be accounted for in a similar manner.  
For $L>7$, the discrepancies are more significant and mostly due to 
the problematic orbits $\gamma=\text{4a}$ and $\gamma={1\over2}(^*10\text{b})$ 
that are not well isolated in phase space and pass close to the cusp of the 
cardioid (cf. Table \ref{badpeaks3pp}). These orbits have inaccurate
Gutzwiller amplitudes for reasons explained in Refs.~\cite{Bruus,paper1}.  

\begin{table}
\begin{center} 
\begin{ruledtabular}
\begin{tabular}{cccc} \\
 $L_\Gamma$ & $\gamma_1$ & $\gamma_2$ & $\gamma_3$ \\  \\ \hline  \\ 
$7.5637$ & ${1 \over 2}(^*$2a) & 4a & \\ 
$7.5650$ & ${1 \over 2}(^*$2a) & ${1 \over 2}(^*$10b) & \\ 
7.9975 & ${1 \over 2}(^*$2a) & ${1 \over 2}(^*$2a) & 4a \\
7.9987 & ${1 \over 2}(^*$2a) & ${1 \over 2}(^*$2a) & ${1\over2}(^*$10b) \\
$8.4731$ & ${1 \over 2}(^*$4b) & 4a & \\
$8.4742$ & ${1 \over 2}(^*$4b) & ${1 \over 2}(^*$10b) & \\ 
$8.8011$ & 2a & 4a & \\
$8.8022$ & 2a & ${1 \over 2}(^*$10b) & \\ 
8.8624 & ${1 \over 2}(^*$2a) & ${1 \over 2}(^*$4b) & 4a \\
8.8636 & ${1 \over 2}(^*$2a) & ${1 \over 2}(^*$4b) & ${1\over2}(^*$10b) \\ \\
\end{tabular}
\end{ruledtabular}
\end{center}
\caption{\label{badpeaks3pp} Some of the orbits responsible for 
numerical discrepancies. The first column gives the length of the periodic 
orbit $\Gamma$ in the full three-particle phase space 
while the other columns specify the constituent periodic orbits $\gamma_i$ 
of the one-particle phase space. (Type-2 hetero-orbits involve only two orbits 
since one of the particles is stationary.)}
\end{table}

\subsection{Symmetry decomposition}\label{symdecompforcard}

Due to the identical nature of the particles, the eigenstates of $\hat{H}$
can be classified according to the irreps of $S_3$, the
permutation group of three identical particles. 
 Each group element belongs to one of three
classes ($(3,0,0)$,$(1,1,0)$,$(0,0,1)$) based on the cycle structure of
that element. Thus, there are also three irreps. These are
the symmetric (trivial) irrep $A_+$, the antisymmetric
irrep $A_-$ and the two-dimensional mixed-symmetry irrep ${\mathcal{E}}$. 
($S_N$ always possesses
exactly two one-dimensional irreps regardless of the size of $N>1$.) 
The character table for $S_3$ is given below. Numbers in front of
class labels indicate the number of elements in that class.

\subsubsection{Quantum mechanics}

The total three-particle density of states can be decomposed into 
symmetry-reduced
densities $\rho_{\mathcal{I}}(E)$, each belonging to an irrep ${\mathcal{I}}$ 
of $S_3$:
\begin{equation}\label{decomprho}
\rho_3(E)=\rho_+(E)+\rho_-(E)+\rho_{\mathcal{E}}(E).
\end{equation}
Each partial density may be obtained by projection $\rho_{\mathcal{I}}(E) =
{\Tr}(\hat{P}_{\mathcal{I}} \delta(E-\hat{H}))$, where the operator 
$\hat{P}_{\mathcal{I}}$
projects onto the irrep ${\mathcal{I}}$
\footnote{For a discete group $G$, the operator $\hat{P}_{\mathcal{I}}$
that projects onto the irrep ${\mathcal{I}}$ is $\hat{P}_{\mathcal{I}}=
(d_{\mathcal{I}}/|G|) \sum_{g} \chi_{_{\mathcal{I}}}(g)
\hat{U}^{\dagger}_{g}$, where the sum is over all group elements $g
\in G$, $d_{\mathcal{I}}$ is the dimension of the
irrep, $|G|$ is the order of the group, $\chi_{_{\mathcal{I}}}(g)$ is the
character of the group element $g$ in the irrep ${\mathcal{I}}$ and 
$\hat{U}_{g}$ is the operator that transforms $\Psi$ as prescribed
by the group element $g \in G$. Permutation operators are unitary
($S_N$ is a unitary group). For $A_{\pm}$ irreps, this reduces to the
operator of Eq.~(\ref{genproj}) and for the irrep ${\mathcal{E}}$ 
of $S_3$, using
the information provided in Table \ref{CtforS3}, $\hat{P}_{\mathcal{E}}=
{2\over6}(2\hat{I}-\hat{U}_{\tau_1}-\hat{U}_{\tau_2})$, where 
$\tau_{1/2} \in (0,0,1)$ and $\hat{I}$ denotes the identity operator.}.  
Expressing the trace in the energy
eigenbasis as in (\ref{sym2pden}), the symmetry-reduced densities are 
\begin{equation}\label{symrhos}
\rho_{\pm}(E)={1 \over 6} \left[ \rho_3(E) \pm  {3 \over 2} \rho_1\left({E
\over 2} \right)*\rho_1(E) +  {2 \over 3} \rho_1\left({E \over
3}\right)\right],
\end{equation}
\begin{eqnarray}\label{rhomixed}
\rho_{\mathcal{E}}(E)={2 \over 6} \left[ 2 \rho_3(E) - {2 \over 3}
\rho_1 \left({E \over 3}\right) \right].
\end{eqnarray}
To understand how the cross term arises in Eq.~(\ref{symrhos}), 
consider the contribution from $\tau = (a)(bc) \in (1,1,0)$:
\bea \label{cross3ppsym} 
\sum_{i,j,k}\braopket{i j k}{\hat{U}_\tau\delta(E-\hat{H})}{i j k}
&=&\sum_{i,j,k}\braket{i k j}{i j k}\delta(E-E_{ijk}) \nonumber \\
&=&\sum_{i,j}\delta(E-E_{ijj}) \nonumber \\  
&=&\sum_{i,j}\delta(E-(\varepsilon_i+2\varepsilon_j)) \nonumber \\ 
&=& {1 \over 2} \rho_1 \left({E \over 2} \right) * \rho_1(E).
\eea 
The other contributions can be found in a similar manner. 
We could compute each partial density separately for comparison with
the numerics, but it is more illuminating to isolate the contribution
from each symmetry class by inverting the above system of equations:
(we ignore the identity class which reproduces the total density of states)
\begin{eqnarray}\label{qnexchngtrms}
\rho_{(1,1,0)}(E) & \equiv & \rho_+(E)-\rho_-(E) \nonumber \\ 
& = & {1\over2}\rho_1\left({E\over2}\right)*\rho_1(E),
\end{eqnarray}
\begin{eqnarray}\label{longliveJamal}
\rho_{(0,0,1)}(E) & \equiv & \rho_+(E)+\rho_-(E)-{1\over2} \rho_{\mathcal{E}}(E) 
\nonumber \\ 
& = & {1\over3}\rho_1\left({E\over3}\right),
\end{eqnarray}
where $\rho_{(1,1,0)}(E)$ and $\rho_{(0,0,1)}(E)$ denote the 
densities belonging to the class of two-particle and
three-particle exchanges, respectively. From Eq.~(\ref{longliveJamal}), we
note that the contribution of the longest cycle is directly related to
the one-particle density of states as discussed at the end of section
\ref{dyncycles}. In the following sections, we
discuss the semiclassical decomposition of each partial density into
smooth and oscillatory components: 
\bea \label{symsmcldecmp}
\rho^{\text{sc}}_{\mathcal{I}}(E)=\bar{\rho}_{\mathcal{I}}(E) + 
\tilde{\rho}_{\mathcal{I}}(E).
\eea

\begin{table}
\begin{center}
\begin{ruledtabular} 
\begin{tabular}{c||ccc}  & & & \\
\hspace*{0.25cm} $S_3$ \hspace*{0.25cm} & $1(3,0,0)$&$3(1,1,0)$&$2(0,0,1)$ \\ 
& & & \\ \hline & & & \\
\hspace*{0.25cm} $A_+$ \hspace*{0.25cm} & 1 & 1 & 1 \\
\hspace*{0.25cm} $A_-$ \hspace*{0.25cm} & 1 & -1 & 1 \\
\hspace*{0.25cm} ${\mathcal{E}}$ \hspace*{0.25cm} & 2 & 0 & -1 \\ & & & \\ 
\end{tabular}
\end{ruledtabular}
\end{center}
\caption{\label{CtforS3} Character table for $S_3$.}
\end{table}

\subsubsection{Stationary cycles}

If all particles being permuted are fixed, the cycles are stationary
and contribute to $\bar{\rho}_{\mathcal{I}}(E)$.
Using the results of Appendix \ref{TFNpsym}, it can be shown that the smooth
densities belonging to each irrep are given by 
Eqs.~(\ref{symrhos}-\ref{rhomixed}), but with $\rho$ replaced by
$\bar{\rho}$. Thus,
\begin{eqnarray}\label{xchngsmth2}
\bar{\rho}_{(1,1,0)}(E) & = &
{1\over2}\bar{\rho}_1\left({E\over2}\right)*\bar{\rho}_1(E) \nonumber \\
& = & {1\over2}\left[{\alpha^2{\mathcal{A}}^2 \over 16 \pi^2}E -
{\alpha^{3/2}{\mathcal{A}}{\mathcal{L}}(1+\sqrt{2})\over 16 \pi^2}\sqrt{E} 
\right. \nonumber \\ 
& + & \left. {\alpha \over 4
\pi}\left({\sqrt{2} {\mathcal{L}}^2 \over 16}+ 3{\mathcal{A}}{\mathcal{K}}
\right)\right], 
\end{eqnarray}
\begin{eqnarray}\label{xchngsmth3}
\bar{\rho}_{(0,0,1)}(E) & = &{1\over3}\bar{\rho}_1\left({E\over3}\right) 
\nonumber \\ 
& = & {1\over 3}\left[{\alpha {\mathcal{A}} \over 4\pi}  -
{\sqrt{3 \alpha} \over 8\pi} {{\mathcal{L}}\over
\sqrt{E}}+3{\mathcal{K}} \delta (E)\right].
\end{eqnarray}
Note that in Eq.~(\ref{xchngsmth2}), we have ignored terms that are
$O(1/\hbar)$ since some of the contributions at this order cannot be
calculated exactly. These terms can be computed numerically, but are
insignificant for our analysis. 

\subsubsection{Hetero-cycles}
The leading-order cycles are the three 1-cycles of the identity class.
If one 1-cycle is stationary and the other two 1-cycles are dynamical
($s=1,e=2$), then the result from 
Eq.~(\ref{thelastbill}) is identical to $\tilde{\rho}^{\h}_{\Gamma}(2,3,E)$.
If instead two 1-cycles are stationary and one 1-cycle is dynamical 
($s=2,e=1$), then 
Eq.~(\ref{thelastbill}) reduces to $\tilde{\rho}^{\h}_{\Gamma}(1,3,E)$.   
There are three contributions of each type. 

The first correction is from permutations $\tau \in (1,1,0)$ that
consist of one 1-cycle and one 2-cycle. There are two such
contributions. The first one is from hetero-cycles for which the
1-cycle is stationary and the 2-cycle is dynamical.
Using formula (\ref{thelastbill}) such that $k=1$ is the
1-cycle and $k=2$ is the 2-cycle ($n_1=1,n_2=2,J_2=2E_2=H=E
\Rightarrow E_2=E/2$), the result has the structure of the
leading-order term of ${1\over2}\tilde{\rho}_1(E/2)*\bar{\rho}_1(E)$.
There is also the situation for which the 2-cycle is stationary and
the 1-cycle is dynamical. Using Eq.~(\ref{thelastbill}) such that
$k=1$ is the dynamical cycle (just a standard periodic orbit of the
one-particle phase space) and $k=2$ is the 2-cycle, the result has the
structure of the leading-order term of ${1\over2}\bar{\rho}_1(E/2)*
\tilde{\rho}_1(E)$.
Group elements $\tau \in (0,0,1)$ consist of single 3-cycles. Therefore,
there are no contributions from this class. To summarise, we have shown that 
\begin{multline}\label{symhetpm}
\tilde{\rho}^{\h}_{\pm}(E)={1\over6}\left[3 \tilde{\rho}^{\text{h1}}_3(E)+3 \tilde{\rho}^{\text{h2}}_3(E) \right. \\
\left. \pm 3
\left({1\over2}\tilde{\rho}_1\left({E\over2}\right)*\bar{\rho}_1(E)+
{1\over2}\bar{\rho}_1\left({E\over2}\right)*\tilde{\rho}_1(E)\right)\right], 
\end{multline}
\begin{eqnarray}\label{symhetE}
\tilde{\rho}^{\h}_{\mathcal{E}}(E)=2 \left[
\tilde{\rho}^{\text{h1}}_3(E)+\tilde{\rho}^{\text{h2}}_3(E)\right].
\end{eqnarray}
   
\subsubsection{Dynamical cycles}

The leading-order contribution to $\tilde{\rho}^{\d}_{\pm}(E)$ 
comes from the identity element $\iota=(a)(b)(c)$ which
consists of three 1-cycles ($m_{\tau}=3; J_1=h_a, J_2=h_b,
J_3=h_c$; $\sum_{k}J_k=H$). Thus, there are 2 independent
commuting generators other than $H$ and so we expect periodic orbit families
of dimension 2. Using
Eq.~(\ref{fullsymallevolve}) and the fact that 1-cycles are
equivalent to periodic orbits of the
one-particle phase space, we find the leading-order term of
$\tilde{\rho}^{\d}_{\pm}(E)$ is $\tilde{\rho}^{\d}_3(E)/6$.

The next contribution is from permutations $\tau \in
(1,1,0)$. There are three elements in this class each consisting of one
1-cycle and one 2-cycle ($m_{\tau}=2$; $k=1, n_1=1$;
$k=2, n_2=2$). Then, for $\tau=(ab)(c)$, $J_1=h_c, J_2=h_a+h_b$ and
similarly for the other elements in this class.  Thus, there is
only one independent generator (other than $H$) and we expect
one-dimensional families. Using Eq.~(\ref{fullsymallevolve}), we find
this contribution has the structure of a two-particle density.
The 1-cycles ($k=1$) are assigned to $\gamma_1$ and the 2-cycles
($k=2$) to $\gamma_2$, where $\gamma_{1/2}$ are any periodic orbits of the
one-particle phase space. Then, all cycle properties are those of the
corresponding orbit (cf. the
1-cycle and 2-cycle of Fig.~\ref{cycle123}; note the repetition 
$l_2/n_2=2/2=1$ which denotes the case where
the particles of the 2-cycle evolve together is not shown). Multiple
repetitions of the 2-cycle are either fractions (if $l_2$ is odd) and
correspond to type-1 DPPOs \emph{or} integers (if $l_2$ is even) and are
type-0 DPPOs (cf. the classification used in section \ref{dynppo}). The generators
$J_1=n_1E_1=E_1$ and $J_2=n_2E_2=2E_2$ are the energies of the
particles involved in the 1-cycle and 2-cycle, respectively (particles
of the 2-cycle \emph{each} have energy $E/2$). Thus, the
final form is structurally equivalent to ${1 \over 2}
\tilde{\rho}_1(E/2) * \tilde{\rho}_1(E)$. 

The two group elements $\tau \in (0,0,1)$ each consist of one
3-cycle ($m_{\tau}=1, k=1, n_1=3$) which implies there are no
generators independent of $H$ and thus the orbits are isolated. 
As before, cycle properties can be mapped to
those of an orbit of the one-particle phase space 
 (cf. the 3-cycle shown Fig.~\ref{uglyfig};  
$l_k/n_k=l_1/n_1=1/3,2/3,3/3$; higher repetitions $l_1/n_1=l_1/3$
would have action $l_1S$,
phase index $l_1 \sigma$, stability matrix $\tilde{M}^{l_1}$, where
$S,\sigma,\tilde{M}$ are the properties of the primitive
orbit to which the cycle is assigned). The
energy $E_1$ in Eq.~(\ref{fullsymallevolve}) is the energy of each
particle involved in the 3-cycle $(k=1)$ and since $H=J_1=3E_1=E$, it
follows that $E_1=E/3$.  Thus, the result has
the structure of a one-particle trace formula, but it is evaluated
at $E/3$ and has a cycle structure prefactor of 1/3.
Including the prefactors from the projection operator, we conclude
that 
\be \label{pmdyn}
\tilde{\rho}^{\d}_{\pm}(E)={1 \over 6} \left[ \tilde{\rho}^{\d}_3(E) 
\pm {3 \over 2}
\tilde{\rho}_1\left({E \over 2}\right) * \tilde{\rho}_1(E)+{2\over3}
\tilde{\rho}_1\left({E \over 3}\right)\right],
\ee
\begin{eqnarray}\label{Edyn}
\tilde{\rho}^{\d}_{\mathcal{E}}(E)={2 \over 6}\left[2 \tilde{\rho}^{\d}_3(E) -
{2\over3}\tilde{\rho}_1\left({E\over3}\right)\right].
\end{eqnarray}
We stress that even though the correction terms have structures
equivalent to one- and two-particle densities, they are in fact
contributions from the dynamical cycles of the full
three-particle phase space. 

\subsubsection{Trace formulas for the two symmetry classes}

Combining the results of Eqs.~(\ref{symhetpm}-\ref{Edyn}), the 
fluctuating densities for the two nontrivial symmetry classes are
\bea \label{oscsym1}
\tilde{\rho}_{(1,1,0)}(E)&=&\tilde{\rho}_{+}(E)-\tilde{\rho}_{-}(E) = 
{1\over2}\tilde{\rho}_1\left({E\over2}\right)*\tilde{\rho}_1(E) 
\nonumber \\ 
&+&\left[{1\over2}\bar{\rho}_1\left({E\over2}\right)*\tilde{\rho}_1(E)+
{1\over2}\tilde{\rho}_1\left({E\over2}\right)*\bar{\rho}_1(E)\right] 
\nonumber \\  
&=& \tilde{\rho}^{\d}_{(1,1,0)}(E) + \left[
\tilde{\rho}^{\text{h1}}_{(1,1,0)}(E) + \tilde{\rho}^{\text{h2}}_{(1,1,0)}(E) 
 \right] \nonumber \\ 
&=&\tilde{\rho}^{\d}_{(1,1,0)}(E) + \tilde{\rho}^{\h}_{(1,1,0)}(E) 
\eea
and
\bea \label{oscsym2}
\tilde{\rho}_{(0,0,1)}(E)& = & \tilde{\rho}_{+}(E)+\tilde{\rho}_{-}(E)
-{1\over2} \tilde{\rho}_{\mathcal{E}}(E) \nonumber \\ 
& = & {1\over3}\tilde{\rho}_1\left({E\over3}\right).
\eea
The leading-order term of $\tilde{\rho}^{\h}_{(1,1,0)}(E)$ is given by  
\begin{equation} \label{sym2het} \begin{split}
& \tilde{\rho}^{\h}_{(1,1,0)}(E) = {{\alpha \mathcal{A}} \over {4 \pi^2}}
\sum_{\gamma} {{({L^0_{\gamma}}/2L_{\gamma})} \over {\sqrt{\left| \det
\left( \tilde{M}_{\gamma} - I \right) \right|}}} \\ 
 & \quad \times \cos\left(\sqrt{\alpha E} L_{\gamma} - \sigma_{\gamma}
{\pi \over 2} -{\pi\over2} \right) \\ 
& \qquad + \left\{ L^0_{\gamma}\rightarrow
\sqrt{2}L^0_{\gamma}, L_{\gamma} \rightarrow {L_{\gamma}\over
\sqrt{2}} \right\}.
\end{split}
\end{equation}
The first term of (\ref{sym2het}) is the contribution from two
particles being stationary at the same point in the billiard (i.e. a
stationary 2-cycle) while the third particle evolves on a periodic
orbit (i.e. a dynamical 1-cycle). The second term is the contribution
from one particle being stationary (i.e. a stationary 1-cycle)
while the other two particles evolve on a periodic orbit
(i.e. a dynamical 2-cycle). 
Higher-order contributions from hetero-cycles can be worked out. 
These are included in the numerics, but we do not write them out
explicitly here (see Appendix \ref{conv3pp}). 
The contribution from the dynamical cycles as
determined above can be written as 
\begin{equation} \label{sym2dyn} \begin{split}
& \tilde{\rho}^{\d}_{(1,1,0)}(E)={\alpha^{3/4}\over({2\pi})^{3/2}E^{1/4}}\sum_{\gamma_1,\gamma_2}
\left(\prod_{i=1}^2{L^0_{\gamma_i} \over
\sqrt{\left|{\det}\left(\tilde{M}_{\gamma_i}-I\right)\right|}}\right) \\
& \quad \times \left[2(2 L^2_{\gamma_1}+L^2_{\gamma_2})\right]^{-1/4}
\cos \left[{\sqrt{\alpha E}\over
\sqrt{2}} L_{12}-\sigma_{12}{\pi\over2}-{\pi\over4}\right],
\end{split}
\end{equation}
where $L_{12}=\sqrt{2L^2_{\gamma_1}+L^2_{\gamma_2}}$ and
$\sigma_{12}=(\sigma_{\gamma_1}+\sigma_{\gamma_2})$. 
To understand how this result is obtained, 
recall the structure of the dynamical
cycles in this class. Each full cycle consists of one 1-cycle and one 2-cycle. 
The total energy is partitioned among the three particles such that 
the periods of the cycles are the same. Suppose the 1-cycle and
2-cycle are associated with the orbits $\gamma_1$ and
$\gamma_2$, respectively.
The energies $E_1,E_2$ are determined from the periodicity condition
\bea \label{statphcon}
T_{\gamma_1}(E_1)={1\over2}T_{\gamma_2}\left({E_2}\right)
\equiv T.
\eea
Using the usual relations for actions and periods in a billiard
(\ref{Tbill}), one can show that 
\be \label{statenergs}
E_1=\left[{2L^2_{\gamma_1}\over{2L^2_{\gamma_1}+L^2_{\gamma_2}}}\right]E, \quad
E_2=\left[{(L^2_{\gamma_2}/2)\over{2L^2_{\gamma_1}+L^2_{\gamma_2}}}\right]E,
\ee
where $E_1$ is the energy of the particle of the 1-cycle, 
$2E_2$ is the total energy of the particles involved in the 2-cycle
(each of them has energy ${E_2}$ since their energies must be equal)
and $E=E_1+2E_2$ is the
total energy of the three-particle system. The 2-cycles are
similar to the DPPOs of a two-particle system (cf. section \ref{dynppo}) 
and we shall use
the same classification scheme for all 2-cycles. 
The trace formula
for $\tilde{\rho}_{(0,0,1)}(E)$ is a one-particle trace formula except
that $L^0_{\gamma}\rightarrow \sqrt{3}L^0_{\gamma}$ and $L_{\gamma} 
\rightarrow {L_{\gamma}\over\sqrt{3}}$.
 
\subsubsection{Numerics}

\begin{figure}
\hspace*{-0.3in} 
\scalebox{0.511}{\includegraphics{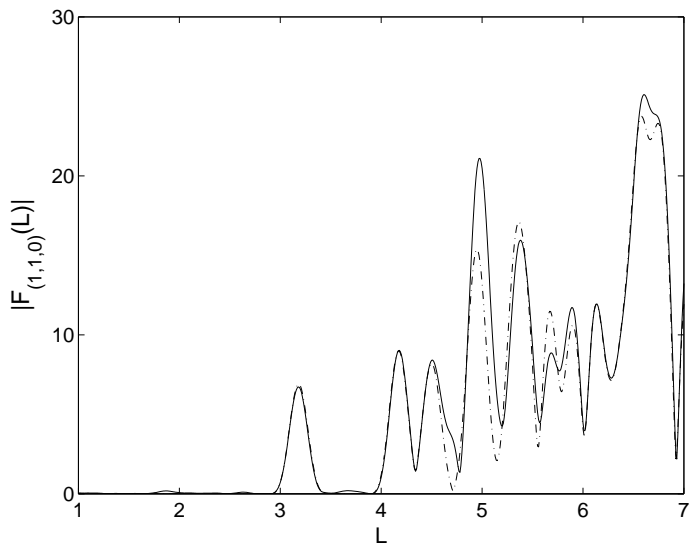}}
\caption{\label{qn2plot} Length spectrum for the class $(1,1,0)$. 
Quantum (solid line) and semiclassical 
(dashed-dotted line) results for $L < 7$. Each peak is due to a
dynamical cycle of the full three-particle phase space.} 
\end{figure}

We first consider the class $(1,1,0)$. We compare numerically the length
spectrum of the dynamical cycles 
\be \label{symqn2dyn}
\tilde{F}^{\text{sc}}_{(1,1,0)}(L)={\mathcal{F}}\{\tilde{\rho}^{\d}_{(1,1,0)}(k)\}
\ee
with its quantum analogue
\be \label{symqn2}
\tilde{F}^{\text{qm}}_{(1,1,0)}(L)={\mathcal{F}}\{\rho_{(1,1,0)}(k)-\bar{\rho}_{(1,1,0)}(k)-\tilde{\rho}^{\h}_{(1,1,0)}(k)\}.
\ee
We construct the first 241,080 levels of $\rho_{(1,1,0)}(k)$ using the
first 1000 single-particle energies. The smooth term is computed from
Eq.~(\ref{xchngsmth2}) using the billiard parameters (as above) for the odd
spectrum. Trace formulas are computed using 
geometrical orbits with length $L<10$. The result is shown in
Fig.~\ref{qn2plot}. 

\begin{figure}
\hspace*{-0.3in} 
\scalebox{0.466}[0.525]{\includegraphics{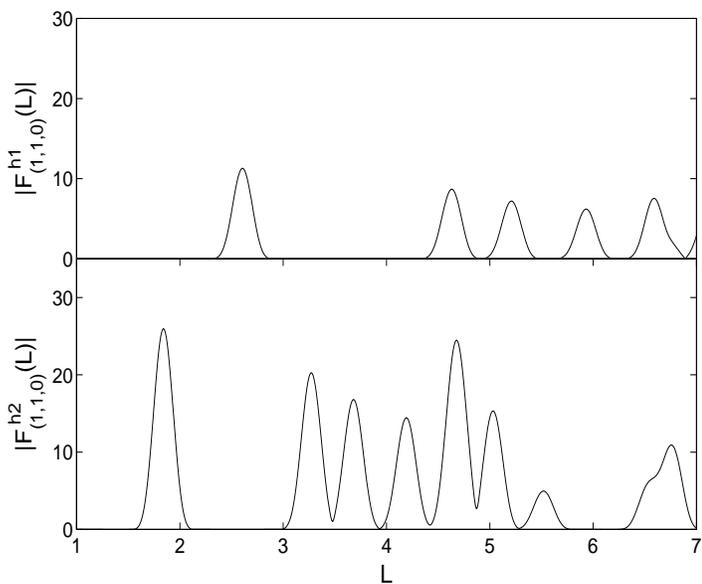}}
\caption{\label{qn2hplot} Length spectrum of the hetero-cycles for the class
$(1,1,0)$. The upper and lower windows show 
${\mathcal{F}} \left\{ \tilde{\rho}^{\text{h1}}_{(1,1,0)}(k)\right\}$ and 
${\mathcal{F}} \left\{ \tilde{\rho}^{\text{h2}}_{(1,1,0)}(k)\right\}$, 
respectively.} 
\end{figure}

We now examine some of the $L$-space structures of Fig.~\ref{qn2plot}. 
The first peak ($L \approx 3.18$) is due to the dynamical cycle where
both 1- and 2-cycles are on the primitive orbit
$\gamma^0={1\over2}(^*$2a). The particle of the 
1-cycle completes one full motion on the orbit while the particles
of the 2-cycle each traverse half the orbit and are then
exchanged. The second peak ($L \approx 4.17$) occurs because of a
dynamical cycle where the 1- and 2-cycles are on primitve
orbits $\gamma^0_1={1\over2}(^*$2a) and $\gamma^0_2={1\over2}(^*$4b), 
respectively. For the third peak ($L \approx 4.50$), the 1-cycle is as
in the first case, except the 2-cycle is type-$0$. 
Thus, as the particle of the 1-cycle completes one full
motion on $\gamma^0={1\over2}(^*$2a), the particles of the 2-cycle each
traverse the full orbit and are then exchanged. This is summarised in
Table \ref{dyncyclesqn2} where some of the dynamical cycles in this
class are listed. In each case, the energies are divided according to
Eq.~(\ref{statenergs}).  For $L < 7$, there are a total of 27
dynamical cycles. All structures in $L$-space can be accounted for in
a similar manner and can be checked systematically by noting that due
to the energy division between the particles, we expect peaks at
positions
$L=\sqrt{L^2_{\gamma_1}+(n^{\pm}_{\gamma_2}L^0_{\gamma_2})^2/2}$. The
2-cycles are type-0 and type-1 for even ($n^{+}_{\gamma_2}$) and odd
($n^{-}_{\gamma_2}$) integer repetition indices, respectively. 

\begin{table}
\begin{center}
\begin{ruledtabular} 
\begin{tabular}{cccc} \\
$L$ & 2-cycle class & $\gamma_1$ & $\gamma^0_2$
  \\  \\ \hline \\
        $3.18$ & $1$ & ${1 \over 2}(^*$2a) & ${1 \over 2}(^*$2a) \\ 
        $4.17$ & $1$ & ${1 \over 2}(^*$2a) & ${1 \over 2}(^*$4b) \\ 
        $4.50$ & $0$ & ${1 \over 2}(^*$2a) & ${1 \over 2}(^*$2a) \\ 
        $4.93$ & $1$ & ${1 \over 2}(^*$2a) & ${1 \over 2}(^*$6b) \\
        $4.97$ & $1$ & ${1 \over 2}(^*$4b) & ${1 \over 2}(^*$2a) \\
        $5.51$ & $1$ & $(^*$2a) & ${1 \over 2}(^*$2a) \\ 
        $5.90$ & $0$ & ${1 \over 2}(^*$4b) & ${1 \over 2}(^*$2a) \\ 
        $6.09$ & $2(1)$ & ${1 \over 2}(^*$2a) & ${1 \over 2}(^*$2a) \\ 
        $6.14$ & $1$  & $(^*$2a) & ${1 \over 2}(^*$4b)  \\ 
        $6.20$ & $1$ & ${1 \over 2}(^*$6b) & ${1 \over 2}(^*$2a) \\ 
        $6.36$ & $0$ & $(^*$2a) & ${1 \over 2}(^*$2a) \\ \\ 
\end{tabular}
\end{ruledtabular}
\end{center}
\caption{\label{dyncyclesqn2}Some dynamical cycles of the 
three-particle cardioid billiard for the class $(1,1,0)$. The first column 
gives the position of the peak in $L$-space arising from the dynamical cycle 
while the third and fourth columns specify the orbits on which 
the 1- and 2-cycles evolve. The second column 
indicates the type of 2-cycle using the classification scheme of section 
\ref{dynppo}. The dynamical cycle that produces a signal at 
$L \approx 6.09$ has a 
prefactor of 2 with its 2-cycle class indicator to denote that 
it is the first repetition of a type-1 2-cycle. (In this case, each particle 
involved in the 2-cycle traverses \textit{one and one-half} of the primitive 
orbit $\gamma^0_2$ before particle exchange.)}
\end{table}  

The discrepancies between quantum and semiclassical results are due to
the problematic orbits mentioned above. The structure at 
$L \approx 5$ is poorly reproduced due to
the hetero-cycles involving 1-cycles that are stationary and 2-cycles
that are dynamical and evolving on the problematic orbits 
$\gamma^0=4{\text{a}}$ and
$\gamma^0={1\over2}(^*$10b). (The length spectrum of the hetero-cycles
is shown in Fig.~\ref{qn2hplot}.) All other discrepancies are due to purely
dynamical cycles and these are summarised in Table \ref{badpeaksqn2}.

\begin{table}
\begin{center} 
\begin{ruledtabular}
\begin{tabular}{ccc} \\
 $L$ & $\gamma^0_1$  & $\gamma^0_2$ \\  \\ \hline \\
        $5.3868$ & ${1 \over 2}(^*$2a) & ${1 \over 2}(^*$8b) \\ 
        $5.6551$ & ${1 \over 2}(^*$2a) & 4a \\ 
        $5.6560$ & ${1 \over 2}(^*$2a) & ${1 \over 2}(^*$10b) \\ 
        $6.6031$ & ${1 \over 2}(^*$4b) & ${1 \over 2}(^*$8b) \\
        $6.8237$ & ${1 \over 2}(^*$4b) & 4a \\
        $6.8245$ & ${1 \over 2}(^*$4b) & ${1 \over 2}(^*$10b) \\ 
        $6.9217$ & ${1 \over 2}(^*$8b) & ${1 \over 2}(^*$2a) \\ \\
\end{tabular}
\end{ruledtabular}
\end{center}
\caption{\label{badpeaksqn2} Some dynamical cycles of the class
$(1,1,0)$ that are responsible for numerical discrepancies. The first column 
gives the position of the signal in $L$-space arising from the dynamical cycle 
while the second and third columns specify the primitive orbits on which 
the 1- and 2-cycles evolve.}
\end{table}

\begin{figure}
\hspace*{-0.33in} 
\scalebox{0.515}{\includegraphics{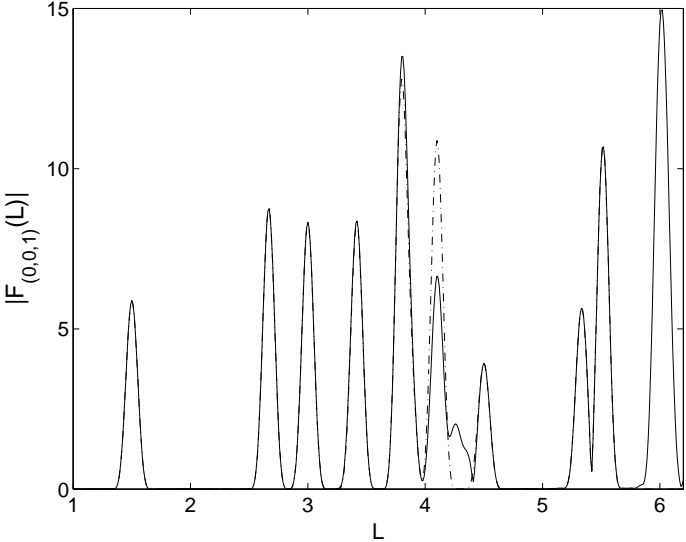}}
\caption{\label{qn3plot} Length spectrum for the class $(0,0,1)$. 
Quantum (solid line) and semiclassical 
(dashed-dotted line) results for $L < 6.25$. Each peak is due to a
dynamical 3-cycle of the full phase space.} 
\end{figure}

We next consider the class $(0,0,1)$. We numerically compare the Fourier
transform of the dynamical 3-cycles 
\bea \label{symqn3dyn}
\tilde{F}^{\text{sc}}_{(0,0,1)}(L)={\mathcal{F}}\{\tilde{\rho}^{\d}_{(0,0,1)}(k)\}
\eea
with its quantum analogue
\bea \label{symqn3}
\tilde{F}^{\text{qm}}_{(0,0,1)}(L)={\mathcal{F}}\{\rho_{(0,0,1)}(k)-\bar{\rho}_{(0,0,1)}(k)\}.
\eea
We use the first 1000 energies of $\rho_{(0,0,1)}(k)$. The smooth term is 
computed from Eq.~(\ref{xchngsmth3}) and
trace formulas are computed using 
geometrical orbits with length $L<11$. The result is shown in
Fig.~\ref{qn3plot}. We now identify some of the peak structures 
 with one or several of the orbits listed in Fig.~\ref{Bruusfig}.

\begin{table}
\begin{center}
\begin{ruledtabular} 
\begin{tabular}{ccc} \\
$L$ & Class & $\gamma^0$ \\  \\ \hline \\
        $1.50$ & $1$ & ${1 \over 2}(^*$2a) \\ 
        $2.67$ & $1$ & ${1 \over 2}(^*$4b)  \\ 
        $3.00$ & $2$ & ${1 \over 2}(^*$2a) \\ 
        $3.42$ & $1$ & ${1 \over 2}(^*$6b) \\
        $3.80$ & $1$ & 3a \\
        $3.85$ & $1$ & ${1 \over 2}(^*$8b) \\ 
        $4.50$ & $0$ & ${1 \over 2}(^*$2a) \\ 
        $5.33$ & $2$ & ${1 \over 2}(^*$4b) \\ 
        $5.52$ & $1$ & ${1 \over 2}(^*$8c) \\ 
        $5.99$ & $1$ & 5a \\ 
        $6.00$ & $2(1)$ & ${1 \over 2}(^*$2a) \\ 
        $6.05$ & $1$ & ${1 \over 2}(^*$10h) \\ \\
\end{tabular}
\end{ruledtabular}
\end{center}
\caption{\label{card3dynqn3}Some dynamical 3-cycles of the 
three-particle cardioid billiard. The first column 
gives the position of the peak in $L$-space arising from the dynamical 3-cycle 
while the third column specifies the primitive orbit on which 
the 3-cycle evolves. The second column 
indicates the type of 3-cycle using the classification scheme of  
Fig.~\ref{uglyfig}. The dynamical 3-cycle that produces a signal at 
$L = 6.00$ has a 
prefactor of 2 with its class indicator to denote that 
it is the first repetition of a type-1 3-cycle. This situation is described 
further in the text.}
\end{table}

The first peak ($L \approx 1.5$) can be identified with a type-$1$ 
dynamical 3-cycle consisting of all three particles evolving on the
orbit $\gamma^0={1\over2}(^*$2a) with the same energy and exactly
$T_\gamma/3$ out of phase and each completing one-third
of the full motion on the orbit and finally being permuted as specified
by $\tau = (acb)$.
The third peak  ($L \approx 3$) is due to a type-$2$ 
dynamical 3-cycle where all three particles evolve on the
orbit $\gamma^0={1\over2}(^*$2a) as above, except that each particle completes
two-thirds of the full motion on the orbit before being 
exchanged  according to $\tau = (abc)$. 
The peak at $L \approx 4.5$ is from a type-$0$ dynamical 3-cycle
consisting of all three particles starting and evolving together in
phase on $\gamma^0={1\over2}(^*$2a), but each completing one full motion on the
orbit and then being trivially exchanged as presribed by any 
group element $\tau \in (0,0,1)$. As an example of a higher
multiple cycle, consider the first repetition of the type-$1$ cycle 
mentioned above. 
It is the same as before except that each particle completes one and
one-third of the motion on the orbit before being permuted. 
This is summarised in Table
\ref{card3dynqn3} where some of the dynamical 3-cycles are listed. 
(The peak at $L \approx 4.25$ is completely undetected by the trace
formula since it arises from a \textit{diffractive}
orbit. Such orbits require a separate analysis since they are not included in
the standard Gutzwiller theory \cite{Bruus}). As before, the discrepancy
that occurs at $L \approx 5$ arises from the two orbits $\gamma^0=4 \text{a}$ and
$\gamma^0={1\over2}(^*$10b).
  
All structures in $L$-space can be accounted for in a similar manner. 
As a systematic check, recall that each dynamical 3-cycle can be
mapped one-to-one with a periodic orbit $\gamma$ of the
one-particle phase space. If the orbit has length
$L_{\gamma}=n_{\gamma}L^0_{\gamma}$, where $n_{\gamma}$ is a repetition
index, it is mapped to a 3-cycle where each particle
executes a fraction $n_{\gamma}/3$ of the full motion on
$\gamma$. We can then write $n_{\gamma}/3=i+{j/3}$, where $i={\text{int}}(n_{\gamma}/3)$ (i.e. integer part of $n_{\gamma}/3$) and ${j/3}$
($j=0,1,2$) is the remainder. If $j \neq 0$, then the orbit with
length $L_{\gamma}$ is
associated with the $i$th repetition of a type-$j$ dynamical
3-cycle. If $j=0$, then it is the $(i-1)$th repetition. To determine
peak positions, we recall that all particles of the 3-cycle have
the same energy $E/3$ and thus we expect peaks at lengths
$L=n_{\gamma}L^0_{\gamma}/\sqrt{3}$. (Recall that
a billiard orbit with length $L_\gamma$ has action
$S_\gamma(\varepsilon)/\hbar=\sqrt{\alpha \varepsilon}L_\gamma$.) 

\section{Conclusion}

We began with the case of two
noninteracting identical particles evolving dynamically on
periodic orbits and explained how the
time-translational symmetry leads to families of periodic
orbits in the full phase space. Using the trace formula for continuous
symmetries \cite{stephen}, we obtained a trace formula 
for the two-particle resolvent
consistent with the dynamical term of the semiclassical two-particle
density of states \cite{paper1}. We also proved identities for the
symmetry-reduced densities of states (see
also Appendix \ref{TF2psymcal}) which were stated without proof in our
previous work \cite{paper1}.  Dynamical pseudo-periodic orbits (DPPOs)
were defined and it was shown that their contribution to the
semiclassical exchange term has the form of a one-particle
trace formula.  We also introduced two-particle
heterogeneous periodic orbits in the full phase space.  We discussed
how the structure of these orbits is different in billiards and in
analytic potentials and the explicit contribution of such orbits to
the two-particle resolvent was determined.

We have also demonstrated that the approach used in this paper yields
results that are consistent with those of the convolution method \cite{paper1}.
In the convolution picture, one is faced with the asymptotic analysis 
of many convolution integrals and the further issue of
spurious contributions from them. The full phase space
formalism, on the other hand, more easily generalises to arbitrary 
particle numbers. It is also
more illuminating since it reveals the underlying structure of the
periodic orbit families. (One useful
property of the convolution approach occurs for billiards where 
significant higher-order corrections from hetero-orbits can be 
explicitly calculated (cf. Appendix \ref{conv3pp})). Most importantly,   
the convolution formalism does not accomodate
interactions and it is necessary to use the full phase space
if interactions are to be included.
The symmetric/antisymmetric resolvent for the case of noninteracting
particles can be expressed as a sum of resolvents, one for each element of
the permutation group. 
As shown above, trace formulas for the oscillatory components can be
written as products over cycles, where each cycle is assigned to a periodic
orbit of the one-particle phase space. The 
results of the paper were applied to a specific problem:  
three noninteracting identical particles in a two-dimensional cardioid
billiard. We found that our semiclassical analysis correctly reproduced the
quantum results and we explained how 
these results could be understood in terms of classical structures in
the full phase space.  

We have assumed that the single-particle dynamics is free of any continuous 
symmetry. If there are additional symmetries, then they must also be
properly accounted for in the theory. If we restrict ourselves to the
noninteracting case, then essentially the only difference from what
we have presented above is that $\mathbf{J}$ and $\mathbf{\Theta}$ become
higher dimensional. (An example would be two particles in a disk billiard.
In this case, there are \emph{four} independent constants of the
motion, any two of $\{E,E_a,E_b\}$ and any two of
$\{L_z,L_{z_a},L_{z_b}\}$. 
Thus, the periodic orbits occur in 3-parameter families.)
One future goal is to consider separately the important zeroth-order problem of
harmonic oscillator potentials. The harmonic oscillator has a higher
degree of symmetry than we are accounting for here ($SU(d)$ in $d$
dimensions).  This project would require using the theory of
Ref.~\cite{nonabelian} which derives a trace formula for systems with more
general symmetries including non-Abelian cases. 

As mentioned above, a major advantage of the formalism presented here    
(as compared to the convolution formalism of Ref.~\cite{paper1})
 is that it can be extended to include
interactions. Any interaction between the particles destroys
the periodic orbit families described above and replaces them with 
a discrete set of
isolated orbits. For weak interactions, we can use a perturbative
method \cite{symbreak} which is applicable to any situation where
continuous symmetries are broken. This program will be explored in a
future publication \cite{usfut}. One could also study (zero-range)
point interactions. Such interactions are often considered as corrections 
to mean-field approximations.  
Semiclassically, point-interactions can be understood
using the formalism of diffractive orbits \cite{diff}.
One can imagine that such an interaction leaves the periodic orbit
families (of the noninteracting system) largely unchanged, but
introduces qualitatively new diffractive orbits.
We plan to explore this scenario in future work.

\begin{acknowledgments}
We thank Rajat Bhaduri, Stephen Creagh, Randy Dumont 
and David Goodings for useful discussions. This work was financially
supported by the Natural Sciences and Engineering Research Council of Canada 
(NSERC).
\end{acknowledgments}

\appendix

\section{Two-particle Thomas-Fermi calculation}\label{TF2pcal}

\noindent 
We first discuss the smooth two-particle density of states and 
its decomposition
into bosonic and fermionic densities. Using the identity 
$\delta(E-h(z_a)-h(z_b))=\int {\d}\varepsilon
\delta(\varepsilon-h(z_a))\delta(E-\varepsilon-h(z_b))$, 
we can show that the leading-order smooth term for
the two-particle density of states is the autoconvolution of the
leading-order smooth term of the one-particle density of
states (\ref{smooth}).
We could verify this term-by-term in the expansion of
$\bar{\rho}_2(E)$, but we can do it more efficiently for all terms as
follows.

We use the partition function $Z(\beta)=
\Tr (\exp{(-\beta\hat{H})})$ which is the Laplace transform 
of the density of states. It is convenient to work with the Wigner
transform which is defined for an arbitrary operator $\hat{A}$ as
\be
\wig{\hat{A}}(z) = \int \d x
\left<{q+{x\over2}}\right|{\hat{A}}\left|{q-{x\over 2}}\right> 
\exp\left(-i{p \cdot x \over \hbar}\right),
\ee
in terms of which the trace is
\be \label{trwig}
\Tr\{\hat{A}\} = {1\over (2\pi\hbar)^n}\int\d z\wig{\hat{A}}(z).
\ee
The trace of a product of two (but not more) operators is given by
\be \label{trwig2}
\Tr\{\hat{A}\hat{B}\} = {1\over (2\pi\hbar)^n}\int\d z 
\wig{\hat{A}}(z)\wig{\hat{B}}(z).
\ee
The Wigner transform of the evolution operator,
$\wig{\exp{(-\beta\hat{H})}}(z)$, can be written as an asymptotic
expansion in powers of $\hbar$, the first few terms of which are
typically retained and used as the smooth approximation to the
partition function. Taking the inverse Laplace transform then gives
the smooth density of states. In particular, the leading-order term of
$\wig{\exp{(-\beta\hat{H})}}(z)$ is 
$\exp{(-\beta\wig{\hat{H}}(z))}$, where the Wigner transform of the
quantum Hamiltonian $\wig{\hat{H}}(z)$ is simply the classical
Hamiltonian which we have denoted by $H(z)$. (There are corrections to
this if the Hamiltonian is not of the kinetic plus potential form.) 
The inverse Laplace transform of this expression yields the
leading-order smooth term (\ref{smooth}).

For two independent particles, the full quantum Hamiltonian is 
the sum of one-particle Hamiltonians and since these are functions of
independent phase space variables,
\be \label{mult}
\wig{\exp{(-\beta\hat{H})}}(z) = 
\wig{\exp{(-\beta\hat{h})}}(z_a)
\wig{\exp{(-\beta\hat{h})}}(z_b).
\ee
Thus, the smoothed two-particle partition function is simply the product
of smooth one-particle partition functions. By the Laplace convolution
theorem, this implies that the smoothed two-particle
density of states is the autoconvolution of the smoothed one-particle
density of states. This same argument can be made for the
exact density of states as an alternate proof of Eq.~(\ref{2ppconv}).

\subsection{Symmetrised $2$-particle Thomas-Fermi term}\label{TF2psymcal}

The bosonic and fermionic partition functions are $Z_\pm(\beta)=
\Tr(\hat{P}_\pm\exp{(-\beta\hat{H})})$, where $\hat{P}_\pm$
are the projection operators defined in Eq.~(\ref{projector}). The
leading-order term is just
the two-particle partition funtion. 
The next term $\Tr(\hat{U}\exp{(-\beta\hat{H})})$
 requires slightly more analysis and can be evaluated directly using
Eq.~(\ref{trwig2}). We begin by finding $\wig{\hat{U}}(z)$.

It is shown in Ref.~\cite{pottf} that for a one-particle system with 
a symmetry axis through the coordinate $q$, 
the Wigner transform of the reflection operator is
\be
\wig{\left(\hat{\mathcal{R}}\right)}(z) =
 \pi\hbar\delta(q)\delta(p),
\ee
where $p$ is the momentum conjugate to $q$. We map our problem onto
that one as follows. First, suppose that the one-particle system is
one-dimensional and define the Jacobi coordinates:
$q=q_a-q_b$, $p=(p_a-p_b)/2$, $Q=(q_a+q_b)/2$ and $P=p_a+p_b$. Then,
exchanging $a$ and $b$ is equivalent to reflecting in $q$ so that the
variable $q$ in the above equation is replaced by $q_a-q_b$ and $p$
is replaced by the conjugate momentum $(p_a-p_b)/2$. Then,
$\wig{\hat{U}}(z)=2\pi\hbar\delta(q_a-q_b)\delta(p_a-p_b)$. If the
one-particle system is higher dimensional, then $\hat{U}$ is the
product of one such inversion in every component. All of them are
independent so that the final result is the product of the individual
ones (for the same reason that Eq. (\ref{mult}) is
multiplicative). The final result is
\be \label{WigtransofU}
\wig{\hat{U}}(z) = (2\pi\hbar)^d\delta(z_a-z_b),
\ee
where the delta function represents the product of all $2d$ delta 
functions (two for each
component). Equivalent results can be found in Ref.~\cite{sommer}.

Then,
\begin{equation} \label{xchtermsmth} \begin{split}
& \Tr\left(\hat{U}\exp{(-\beta\hat{H})}\right)  \\
& = {1\over (2\pi\hbar)^{2d}}
\int \d z \wig{\hat{U}}(z) \wig{\exp{(-\beta\hat{H})}}(z) \\
& = {1\over (2\pi\hbar)^{d}}\int \d z_a 
\wig{\exp{(-2\beta\hat{h})}}(z_a), 
\end{split}
\end{equation}
where we used the delta functions from $\wig{\hat{U}}(z)$ in
Eq.~(\ref{WigtransofU}) to do the integrals over the $z_b$ variables
and the multiplicative property of the Wigner functions as in 
Eq.~(\ref{mult}). The first few terms of Eq.~(\ref{xchtermsmth}) give 
the smooth approximation to the one-particle partition
function evaluated at $2\beta$. Under the inverse Laplace transform,
this becomes $\bar{\rho}_1(E/2)/2$ and we conclude
\be \label{divbar}
\bar{\rho}_\pm(E) = {1\over 2}\left(\bar{\rho}_2(E) \pm 
{1\over 2}\bar{\rho}_1\left({E\over2}\right)\right),
\ee
consistent with (\ref{symmreddos}).

\section{Stability matrix of pseudo-periodic orbits}\label{determinants} 
We first prove that
$\det(\tilde{M}_{\gamma'}-I)=4 
\det(\tilde{M}_\gamma-I)$, where $\tilde{M}_{\gamma'}$ is
the stability matrix of a pseudo-periodic orbit ${\gamma'}$ of the
full two-particle phase space and $\tilde{M}_\gamma$ is the stability
matrix of the corresponding periodic orbit $\gamma$ in the one-particle phase
space. This admits various generalisations which are used in the main
discussion. The (type-1) dynamical  
pseudo-periodic orbit (DPPO) consists of both particles evolving
for half of the single-particle period $T_\gamma/2$ followed by the
symplectic mapping $u$ that exchanges the two particles. 

\begin{figure} 
\scalebox{0.3}{\includegraphics{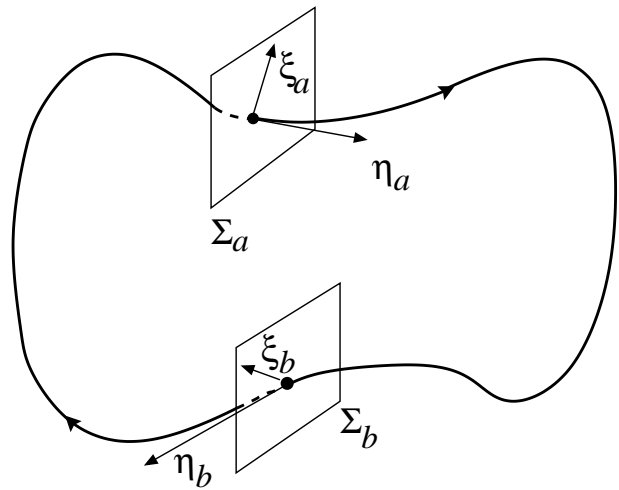}}
\caption{\label{sectionsfig} The coordinates of particles 
$a$ and $b$ on a (type-1) dynamical 
pseudo-periodic orbit ${\gamma'}$ of the full phase space (which
can be mapped one-to-one to an orbit $\gamma$ of the one-particle phase
space). $\Sigma_a$ denotes an initial section for particle $a$, $\eta_a$
is the coordinate along the orbit (the coordinate transverse to the
$h_a$ surface denoted by $\kappa_a$ is not shown) and $\xi_a$ are
the $(2d-2)$ remaining
coordinates for particle $a$ which lie on $\Sigma_a$. (All points on
$\Sigma_a$ are at equal energy.) Similarly, for particle $b$.}
\end{figure}

We define
coordinates as follows (cf. Fig.~\ref{sectionsfig}). For particle
$a$, we define an initial section $\Sigma_a$ such that the phase space
flow is transverse to it and all points on the section are at equal
energy. We define a coordinate pointing along the orbit which we call
$\eta_a$. Without loss of generality, we can take $\partial
\eta_a / \partial t \equiv 1$. We also define a coordinate transverse to the
constant $h_a$ surface (but in the phase space of particle $a$) which
we call $\kappa_a$. If we consider it to take the values of $h_a$,
then it is canonically conjugate to $\eta_a$ and has zero time
derivative under the flow since $h_a$ is conserved.
The remaining $(2d-2)$ coordinates for particle $a$ lie on the section
$\Sigma_a$ and will be collectively denoted by $\xi_a$. As the flow
evolves, changes in the $\xi_a$ coordinates are described by the
$(2d-2) \times (2d-2)$ symplectic stability matrix (for the
one-particle dynamics) $\tilde{N}_a$.  Similarly, we define
$\Sigma_b$, $\eta_b$, $\kappa_b$, $\xi_b$ and $\tilde{N}_b$ for
particle $b$. We also need a way to connect coordinates on $\Sigma_a$
to those on $\Sigma_b$; we will take them to be such that they are
connected by parallel transport so that, for example, the stable and
unstable manifolds are mapped onto each other.

We start by defining the symplectic transformation
\begin{eqnarray}\label{com}
\eta = {\eta_a+\eta_b\over 2},
\qquad
\kappa = \kappa_a+\kappa_b, \nonumber \\
\upsilon = \eta_a-\eta_b,
\qquad
\zeta = {\kappa_a-\kappa_b\over 2}.
\end{eqnarray}
The monodromy matrix $M_{\gamma'}$ describes the linearised motion of
small perturbations around a DPPO ${\gamma'}$ of the
full phase space. In particular, if $\mathbf{\Upsilon}= (\kappa, \eta,
\upsilon, \zeta, \xi_a, \xi_b)$, then $\delta \mathbf{\Upsilon}=
M_{\gamma'} \delta \mathbf{\Upsilon}_0$. Consider an
initial slight change in $\eta$ by the amount $\delta\eta_0$ while
keeping all other coordinates constant. This implies that both
$\eta_a$ and $\eta_b$ increase by $\delta\eta_0$. After time evolution
for $T_\gamma/2$ and particle exchange, $\delta\eta=\delta\eta_0$ while
all other coordinates are unchanged (in particular, the transverse
coordinates are unaffected). Now consider an initial small change in
$\kappa$ by the amount $\delta\kappa_0$. This implies that $\kappa_a$
and $\kappa_b$ change by $\delta\kappa_0/2$. After integrating for
time $T_\gamma/2$ and interchanging the particles, we observe that
$\delta\kappa=\delta\kappa_0$. However, this change of value in
$\kappa$ does affect the $\eta$ coordinate. Under this change, the
period of the orbit $\gamma$ also changes; let $T_\gamma'$
denote the derivative of this period with respect to the
single-particle energy. Since we are only integrating for half of the
period, and the single-particle energies are changed by
$\delta\kappa_0/2$, we find that $\delta\eta =
-T_\gamma'\delta\kappa_0/4$. (The minus sign indicates that if the
period increases and we integrate for the same amount of time as
before, then the particles will fail to execute a complete loop,
corresponding to a negative value of $\eta$.) Thus, the monodromy
matrix of the pseudo-periodic orbit $\gamma'$ in the full phase space
has the form
\begin{equation}
M_{\gamma'}= \left(
\begin{array}{ccc}
   \begin{array}{cc}
         1 & 0  \\
       -{T_\gamma'\over 4} & 1
   \end{array} & {\mathbf 0} \\
{\mathbf 0} & \tilde{M}_{\gamma'}
\end{array}
\right).
\end{equation}
We are interested in calculating $\det(\tilde{M}_{\gamma'}-I)$
to evaluate the Gutzwiller amplitude. Note that the matrix
$\tilde{M}_{\gamma'}$ involves only the $(4d-2)$ phase space
coordinates other than $\eta$ and $\kappa$.
We can understand the calculation up to now as follows. The
transformation (\ref{com}) can be thought of as a transformation to
center-of-mass coordinates. We have removed the center-of-mass
coordinates $\eta$ and $\kappa$ from consideration and are only left
with the relative coordinates $\upsilon$ and $\zeta$ (as well as all
the transverse coordinates $\xi_a$ and $\xi_b$.) It is 
reasonable that only the relative coordinates are important for
determining the stability.

The next two coordinates we consider are $\upsilon$
and $\zeta$. Let us start with $\upsilon$. A small initial change in
$\upsilon$ by the amount 
$\delta\upsilon_0$ implies
that $\eta_a$ changes by $\delta\upsilon_0/2$ while $\eta_b$ changes by
$-\delta\upsilon_0/2$. After integrating for time $T_\gamma/2$, this remains
unchanged, but after particle exchange, the final values of $\delta\eta_a$
and $\delta\eta_b$ are changed in sign so that the corresponding
diagonal matrix element of $\tilde{M}_{\gamma'}$ is $-1$. Similarly, the
diagonal matrix element corresponding to the $\zeta$
coordinate is also $-1$. As before, an infinitesmal change in
$\zeta$ implies an
infinitesmal change in $\upsilon$. In this case, the corresponding matrix
element is $T_{\gamma}'$. Therefore, we can write
\begin{equation} \label{aaa}
\tilde{M}_{\gamma'}= \left(
\begin{array}{ccc}
   \begin{array}{cc}
   -1 & 0 \\
   T_\gamma' & -1
   \end{array} & {\mathbf 0} \\
{\mathbf 0} & \tilde{N}
\end{array}
\right).
\end{equation}
Then, $\det(\tilde{M}_{\gamma'}-I) = 4 \det(\tilde{N}-I)$, where
we use appropriately dimensioned identity matrices on each side of the
equality. It remains to calculate the determinant of the
$(4d-4)\times(4d-4)$ matrix $\tilde{N}-I$. The matrix $\tilde{N}$
involves only the coordinates $\xi_a$ and $\xi_b$. Since these two
sets of coordinates live on different sections, we cannot immediately
define a mapping between them. To do so, we note that we have defined
coordinates on the two sections so that the exchange operation is a
simple mapping of the form
\begin{equation}
\tilde{E}=\left(
\begin{array}{cc}
{\mathbf 0} & I\\
I & {\mathbf 0}
\end{array}
\right),
\end{equation}
where $I$ is a $(2d-2) \times (2d-2)$ identity matrix. In terms of
these coordinates, the one-particle stability matrices $\tilde{N}_a$
and $\tilde{N}_b$ are such
that $\tilde{N}_a \tilde{N}_b=\tilde{M}_\gamma$, which is the stability
matrix of the full periodic orbit for the one-particle
dynamics. The combined operations of flow and exchange give
\begin{equation}
\tilde{N}=\left(
\begin{array}{cc}
{\mathbf 0} & \tilde{N}_b \\
\tilde{N}_a & {\mathbf 0}
\end{array}
\right)
\end{equation}
and
\begin{eqnarray}\label{simplifyNtilde}
\det(\tilde{N}-I) & = &
\det \left(
\begin{array}{cc}
-I & \tilde{N}_b\\
\tilde{N}_a & -I
\end{array}
\right) \nonumber \\
&  = & 
\det \left(
\begin{array}{cc}
\tilde{N}_a & -I \\
-I  & \tilde{N}_b
\end{array}
\right) \nonumber \\
& = &
\det(\tilde{M}_\gamma-I),
\end{eqnarray}
where after the second equality, we interchanged rows to put the matrix into a
more useful form. (The matrix has even dimension so there is no sign
introduced as a result of this interchange.) The final equality of
Eq. (\ref{simplifyNtilde}) requires the following identity.  If a
matrix $C$ has the form
\begin{equation}
C=\left(
\begin{array}{cc}
A & -I\\
-I & B
\end{array}
\right),
\end{equation}
then $\det (C) = \det (AB-I)$. 
This can be shown by multiplying $C$ by the matrix 
\begin{equation} C' =
\left(
\begin{array} {cc}
B & I\\
I & A
\end{array}
\right).
\end{equation}
After multiplying them together, the product is block-diagonal with
$(AB-I)$ in one block and $(BA-I)$ in the other. These have equal
determinants. Since $C'$ has the same determinant as $C$, it follows
that $[\det (C)]^2=[\det (AB-I)]^2$ and thus we have
identified the two determinants within a sign. The sign follows from
the fact that the contribution to the determinant of the fully
diagonal term $\prod A_{ii}B_{ii}$ should be positive. Thus, we
conclude that $\det(\tilde{M}_{\gamma'}-I) = 4 
\det(\tilde{M}_\gamma-I)$. 

It is a straightforward extension to generalise this result to a cycle
with $n$ particles on an orbit. We first have to find some
appropriate set of variables so that we may
isolate an upper-left block of the monodromy matrix in analogy to 
Eq.~(\ref{aaa}). This comes from the $2n$ coordinates $\eta$ and
$\kappa$. As in the previous case, we separate the variables into
center-of-mass coordinates (which subsequently play no role) and a set
of relative (Jacobi) coordinates. Using similar arguments to the $n=2$
case, the determinant of the upper-left
block is then $n^2$. The contribution from the rest of the matrix
(\textit{i.e.} the lower-right block) comes from the transverse
coordinates. In terms of these transverse coordinates, the
single-particle stability matrices
$\tilde{N}_a, \tilde{N}_b, \ldots, \tilde{N}_n$ are such that
$\tilde{N}_a\tilde{N}_b\cdots\tilde{N}_n=\tilde{M}_\gamma$, which is
the stability matrix of the full periodic orbit for the one-particle
dynamics. Through a sequence 
of manipulations and transpositions similar to the $n=2$ case, the 
determinant of this lower-right block 
(\textit{i.e.} $\det(\tilde{N}-I)$) can then be reduced to the form 
\be
\det \left(
\begin{array}{cccccc}
\tilde{N}_a & -I          &  0     & \cdots & 0              & 0 \\
0           & \tilde{N}_b & -I     & \cdots & 0              & 0 \\
\vdots      & \vdots      & \vdots & \ddots & \vdots         & \vdots \\
 0          & 0           & 0      & \cdots &\tilde{N}_{n-1} & -I \\
-I          & 0           &  0     & \cdots & 0              & \tilde{N}_n 
\end{array}
\right),
\ee
which is a generalisation of the $n=2$ case.  
It can be shown that $\det(\tilde{N}-I)=\det(\tilde{M}_\gamma-I)$ and 
thus, $\det(\tilde{M}_{\gamma'}-I) = n^2  
\det(\tilde{M}_\gamma-I)$. 
If the orbit is not primitive, but is a repetition of some simpler
orbit, then we can absorb this into the definitions
of the single-particle stability matrices and carry through all of the 
manipulations as before. The result is unchanged.

\section{Monodromy matrix and phase indices of a harmonic oscillator}
\label{appharmonic} 
It is shown in Ref.~\cite{bb} that the monodromy matrix for
a primitive orbit along the $x$-axis is
\begin{eqnarray}\label{monoharmono}
M =\left( \begin{array}{cc}
     \cos(\omega_yT) & {1 \over \omega_y} \sin(\omega_yT) \\
    -\omega_y \sin(\omega_yT) & \cos(\omega_yT) \end{array} \right),
\end{eqnarray}
where $\omega_y$ is the frequency of the $y$-motion and
$T=2\pi/\omega_x$ is the period of the $x$-motion. This is 
derived by integrating the harmonic oscillator equations of motion for 
time $T$. Then, $\det(M-I)=2\sin(\omega_yT/2)$. In higher
dimensions, the monodromy matrix is simply block diagonal so that
$\det(M-I)=\prod_j 2\sin(\omega_jT/2)$, where the product is over
the directions other than $x$. 

The only role played by
the $x$ variable above was to specify the time of evolution in
the determination of the arguments of the sinusoids. It was not
important that it be a harmonic motion, it is enough that it be
periodic. These results apply for any periodic orbit with period $T$
as long as the transverse motion is harmonic. This exactly describes a
heterogeneous orbit. This justifies the amplitude factor in the
denominator of Eq.~(\ref{scont}) for $d=1$. In higher dimensions, the
stability is given by both the single-particle monodromy matrix of the
evolving particle and by the harmonic motion of the stationary
particle about its potential minimum. But these motions are uncoupled
and so are simply multiplicative in their combined contribution to the
amplitude.

The phase factor can be determined in an analogous manner from 
the exact harmonic oscillator trace formula. For the primitive orbit
along the $x$-axis, the phase index is 3. A factor of 2
arises from the two turning points experienced by the periodic orbit
in traversing the $x$-motion independent of the harmonic motion
transverse to the orbit. The remaining factor of 1 can be
attributed to the harmonic $y$-motion and is related to the sign of
the determinant of the monodromy matrix. For heterogeneous orbits,
this means that we should simply include a phase factor of $-\pi/2$
for the transverse harmonic motion in addition to any phase factors
from the single-particle motion along the periodic orbit. 
In higher dimensions, each
transverse direction is independent and the phase index is 
additive. This accounts for the phase factor of $-d\pi/2$ in
Eq.~(\ref{scont}). The fact that each transverse direction is uncoupled
from all the rest as well as from the single-particle dynamics
transverse to the periodic orbit allows us to simply multiply the
amplitudes and add the phase factors.

Finally, if the potential is a local maximum in one of the directions,
this corresponds to the case of an unstable harmonic
oscillator. It is straightforward to show that its contribution to
$\det(M-I)$ is $2\sinh(\omega_yT/2)$. Furthermore, its phase index is
trivially zero since an unstable periodic orbit running along a ridge
does not fold back on itself and introduces no caustics in phase
space. This fact is also consistent with the trace formula for an
unstable harmonic oscillator as described in Ref.~\cite{bb}.
In higher dimensions with a mixture of stable and unstable directions, 
we continue to multiply the amplitude factors and add the phase
indices of the separate directions. This fully accounts for the
modifications described immediately below Eq.~(\ref{ladedah}).

\section{Symmetrised $N$-particle Thomas-Fermi term}\label{TFNpsym}

We now discuss the smooth contribution to ${\Tr}
(\hat{U}_\tau\hat{G})$ and we shall evaluate it using Wigner
transforms as in Appendix \ref{TF2psymcal}. We need to determine the smooth
approximation to the symmetric/antisymmetric partition function 
\begin{equation} \label{partsym} \begin{split}
& Z_{\pm}(\beta)={\Tr}\left(\hat{P}_{\pm}\exp (-\beta \hat{H})\right) \\
& = {1\over N!} \sum_{\tau}(\pm 1)^{n_{\tau}}{\Tr}
\left(\hat{U}_{\tau}\exp (-\beta \hat{H})\right) \\
& = {1\over N!} \sum_{\tau}(\pm 1)^{n_{\tau}} \int {{\d}z \over
(2\pi\hbar)^{Nd}} \wig{(\hat{U}_\tau)}(z) \wig{\left(\exp (-\beta
\hat{H})\right)}(z).
\end{split}
\end{equation}
Since each group element can be decomposed into independent cycles,
\be
\wig{(\hat{U}_\tau)}(z) = \prod_k f_k(z_k),
\ee
where $k$ indicates the different cycles comprising the group element
$\tau$, $f_k$ is a function we discuss below and $z_k$ denotes the
phase space coordinates of the $n_k$ particles being permuted by that
cycle. (For each group element, the unique decomposition into cycles
also provides a unique decomposition of the phase space into the
subspaces corresponding to the cycles.) The function $f_k$ can be specified
without loss of generality by choosing to label the particles being
permuted by the cycle as $1,2,\ldots,n_k$ (\textit{i.e.} $1 \rightarrow
2$, $2 \rightarrow 3$, $\ldots, n_k
\rightarrow 1$) and to leading order in $\hbar$ \cite{sommer,pottf},  
\be
f_k(z_k) \approx (2\pi\hbar)^{(n_k-1)d}
\delta(z_1-z_2)\delta(z_2-z_3) \cdots \delta(z_{n_k-1}-z_{n_k}). 
\ee

The first group element of the sum in Eq.~(\ref{partsym}) is 
the identity element for
which the decomposition into cycles is the trivial one where each
particle is in a cycle by itself so that all of the $f_k$ are
identically unity. Integrating the smooth approximation to the 
Wigner function of $e^{-\beta\hat{H}}$ yields the smooth $N$-particle
partition function. Using the generalisation of property (\ref{mult}), 
we observe that the leading-order term of $\bar{Z}_{\pm}(\beta)$ 
is just the $N$th power of the
single-particle smooth partition function $\bar{Z}_1(\beta)^N$ and
under the inverse Laplace transform, this is just the 
$(N-1)$-fold convolution integral of the single-particle smooth
density of states.
The prefactor of $1/N!$ comes from the projection operator (\ref{genproj}) and
we conclude that the identity term is $O(1/N!\hbar^{Nd})$. The
first correction will come from group elements that 
consist of one $2$-cycle and $(N-2)$ $1$-cycles. The contribution from
this class will have the form $\bar{Z}_1(\beta)^{N-2}\bar{Z}_1(2\beta)$.  
Compared to the leading-order
term, this class contributes to the density of states with relative
order $O(N\hbar^{d}/2)$. The factor of $N$ is due to the fact that
this class has $N(N-1)/2$ members and they all contribute
identically. The factor of $2$ comes from the inverse Laplace transform
since the argument of one of the single-particle
partition functions is $2\beta$.  The general structure then
emerges. For an arbitrary group element, the contribution to the smooth
partition function is $\prod_k \bar{Z}_1(n_k\beta)$.  It contributes to the
smooth density of states with relative order
$O(\hbar^{(N-m_\tau)d}w_\tau)$, where $m_\tau$ is the number of
independent cycles in the decomposition of $\tau$. The factor $w_\tau$
is the size of the class (a combinatoric factor which
can be found from Eq.~(1-27) of Ref.~\cite{hamermesh}) divided by a
factor arising from the inverse Laplace transform which equals
$\prod_k n_k$.

As a formal expansion in powers of $\hbar$, this may be
inconsistent since some of the neglected corrections from the first few
group elements may be of more significant order than the
leading-order contributions of later group elements. However, for
large $N$, one could easily imagine that the combinatoric factor
$w_\tau$ offsets this effect. Keeping the leading-order $\hbar$ term
of all the group elements then guarantees that one has a good
approximation regardless of the relative size of $1/\hbar$ and $N$.
 
\section{Higher-order $\hbar$ contributions from hetero-orbits in 2-D
billiards}\label{conv3pp}

In this paper, we only discuss leading-order contributions to the
oscillatory part of the density of states. For billiards, 
hetero-orbit families generally have higher dimensionality than
dynamical orbit families and the corrections from the former can
be quite significant. The formalism of Ref.~\cite{stephen} 
cannot be used to determine these corrections. We calculate correction
terms arising from hetero-orbits 
using the convolution method as in Ref.~\cite{paper1}.  
The first few terms of the
asymptotic series can be determined by convolving the Weyl expansion 
(\ref{smooth_bill}) term-by-term with a two-particle trace formula. 
As a formal expansion in powers of $\hbar$, this is
inconsistent since we do not include corrections to the one-particle 
trace formula (\ref{rough}) itself. 
However, our numerics seem to indicate that the 
corrections to the Gutzwiller trace formula are negligible in this case, 
otherwise we could not reproduce the quantum results with such accuracy.  

The contribution from the first type of hetero-orbit where one
particle evolves while the others are stationary is calculated from   
\bea \label{hetconv1}
\tilde{\rho}^{\text{h1}}_3(E)&=&
\bar{\rho}_1(E) * (\bar{\rho}_1 * \tilde{\rho}_1)(E) \nonumber \\
&=&\int_0^E \bar{\rho}_1(\varepsilon) (\bar{\rho}_1
*\tilde{\rho}_1)(E-\varepsilon) {\d} \varepsilon. 
\eea
After convolving Eq.~(\ref{smooth_bill}) with the oscillatory function
$(\bar{\rho}_1 * \tilde{\rho}_1)(E)$ 
(which has been calculated in Ref.~\cite{paper1}), we find 
there are nine integrals to do, but three of these are trivial because
of a delta function in the integrand. The remaining six integrals 
require careful analysis. As an example, we obtain the asymptotic
expansion of  
one integral. The others are calculated in the same
manner, but we forego the details. The first correction to the
leading-order result
(\ref{H1}) comes from two terms; the first (second) is the convolution of
the area (perimeter) term of $\bar{\rho}_1(E)$ with the perimeter (area) 
term of $(\bar{\rho}_1 * \tilde{\rho}_1)(E)$. The integral involved in
the first term is

\be \label{AL}
I_{{\mathcal{A}}{\mathcal{L}}}(E)=\int_0^E(E-\varepsilon)^{-1/4} \cos
\left(a \sqrt{E-\varepsilon}+b \right) {\d} \varepsilon,
\ee
where $a=\sqrt{\alpha}L_\gamma, b=-\sigma_\gamma
{\pi/2}-{\pi/4}$. We want to perform a local analysis about
$\varepsilon=0$. The reason for this is that small $\varepsilon$
corresponds physically to the situation where the stationary particles 
have little energy and most of the energy belongs to the dynamically
evolving particle. For $\varepsilon \approx E$, we have the opposite
situation which does not make sense physically. We first make several changes
of variable to simplify the calculation. 
A first change of variable $u=(E-\varepsilon)^{1/2}$ removes the
square root in the argument of the sinusoid.  
Next, we wish to make a Taylor expansion
about the point  $u=\sqrt{E}$ and to facilitate this, we make a
second change of variable $x=\sqrt{E}-u$. The integral then becomes 
\begin{equation} \label{ALapprox} \begin{split}
&I_{{\mathcal{A}}{\mathcal{L}}} (E)= 2 \int_0^{\sqrt{E}}(\sqrt{E}-x)^{1/2} \cos \left(a
(\sqrt{E}-x)+b \right) {\d} x  \\
& \quad \approx 2 {\text{Re}} \left\{
e^{ia\sqrt{E}+b}\int_0^{\infty}(\sqrt{E}-x)^{1/2}e^{-iax}{\d x}\right\}. 
\end{split}
\end{equation}
The integral 
$\int_0^{\sqrt{E}}(\cdots)=\int_0^{\infty}(\cdots)-\int_{\sqrt{E}}^{\infty}
(\cdots)$; the second integral is an endpoint correction, but
asymptotic in $E$, this correction is negligible. 
Thus, we are justified in 
replacing $\sqrt{E}$ with $\infty$ in the second line
above. At this point, we
obtain the Taylor expansion of $f(x)=(\sqrt{E}-x)^{1/2}$ about $x=0$:
\bea \label{Taylor}
f(x)=E^{1/4}-{1\over2}E^{-1/4}x-{1\over8}E^{-3/4}x^2+\ldots .
\eea 
Since the final correction [$I_{{\mathcal{K}}{\mathcal{L}}}(E)$] 
in the expansion
(\ref{hetconv1}) is $O(E^{-1/4})$, it is only necessary to include terms in
the Taylor series to $O(E^{-1/4})$. Thus, 
\begin{equation}\label{I1series} \begin{split}
 & \int_{0}^{\infty} (\sqrt{E}-x)^{1/2}  e^{-iax}{\d}x \\ 
& \approx E^{1/4} \int_{0}^{\infty} e^{-iax}{\d}x - {1\over2}E^{-1/4}
\int_{0}^{\infty}x e^{-iax}{\d}x \\
& =   E^{1/4} \left(-{i \over a}\right) - {1\over2}E^{-1/4} 
\left(-{1 \over {a^2}}\right)
\end{split}
\end{equation}
and asymptotically, 
\begin{equation}\label{I1exp} \begin{split}
   I_{{\mathcal{A}}{\mathcal{L}}}(E) & \sim {2 \over a} \left\{ E^{1/4} \cos
   \left(a\sqrt{E} + b -{\pi\over2}\right) \right. \\ 
& \left. + {E^{-1/4}\over2a} \cos\left(a\sqrt{E} + b\right)\right\}.
\end{split}
\end{equation}  
An equivalent approach is to evaluate the integral exactly and then replace
the resulting functions with their asymptotic forms. 
Evaluating the integral (\ref{AL}) at the upper limit using this
method then corresponds to the situation where one of the stationary particles 
has all of the energy while the dynamically evolving particle has no
energy. Physically, this does not make sense and this is evident
mathematically since the contribution that comes from evaluating the
integral at the upper endpoint is a smooth function of $E$ and is therefore 
spurious in the sense discussed in Refs.~\cite{spurious,paper1}. The result is
only meaningful if we drop this smooth contribution. This is justified
since we know that any smooth contribution to the density of
states is already contained in the $\bar{\rho}_3(E)$ term. This is
completely equivalent to what is done above (the spurious smooth
contribution above is the endpoint correction that was dropped in the
second line of Eq.~(\ref{ALapprox})). All six convolution 
integrals can be analysed in
this way. Collecting the contributions from all six integrals, 
the expansion up to $O(1/\hbar^{3/2})$ is  
\begin{widetext}
\begin{equation} \label{h1convfin} \begin{split}
\tilde{\rho}^{\text{h1}}_3(E) & = \sum_{\gamma}
{L^0_{\gamma}\over \sqrt{\left |\det 
\left( \tilde{M}_{\gamma}-I \right) \right |}}
 \left[ -{\alpha^{3/2}{\mathcal{A}}^2E^{1/2} \over 8 \pi^3
L^2_\gamma}  \cos \left(\Phi \right) 
-{\alpha^{5/4}{\mathcal{A}}{\mathcal{L}}E^{1/4} \over 8
\sqrt{2}\pi^{5/2}L^{3/2}_\gamma} \cos \left(\Phi  - {3 \pi \over
4}\right) \right. \\
& + \left. {\alpha \over 2 \pi^2 L_\gamma} \left({{\mathcal{A}}^2
\over 4 \pi L^2_\gamma}+ {{\mathcal{L}}^2 \over 32} + {\mathcal{A}}
{\mathcal{K}} \right) \cos \left( \Phi - {\pi\over2}
\right) - {\alpha^{3/4}{\mathcal{L}}E^{-1/4} \over 8 \sqrt{2}
\pi^{3/2}L^{1/2}_\gamma} \left({3 {\mathcal{A}}\over 8 \pi L^2_\gamma} +
{\mathcal{K}}\right) \cos \left(\Phi - {\pi\over4} \right) \right],
\end{split}
\end{equation}
\end{widetext}
where $\Phi=\sqrt{\alpha E}L_\gamma - \sigma_\gamma {\pi/ 2}$.
The contribution from the second type of hetero-orbit where one
particle is stationary while the others evolve is calculated from   
\bea \label{hetconv2}
\tilde{\rho}^{\text{h2}}_3(E) & = & \bar{\rho}_1(E) * (\tilde{\rho}_1 *
\tilde{\rho}_1)(E) \nonumber \\ 
& = & \int_0^E \bar{\rho}_1(\varepsilon) (\tilde{\rho}_1
*\tilde{\rho}_1)(E-\varepsilon) {\d} \varepsilon. 
\eea
After convolving Eq.~(\ref{smooth_bill}) with the formula for 
$(\tilde{\rho}_1 *
\tilde{\rho}_1)(E)$ (which has been calculated in Ref.~\cite{paper1} 
using stationary phase asymptotics), we find there are only two
convolution 
integrals which require analysis. These are evaluated asymptotically 
using the same
technique as above. The final result (up to $O(1/\hbar^{3/2})$) is
\begin{widetext}
\begin{equation} \label{h2convfin} \begin{split}
\tilde{\rho}^{\text{h2}}_3(E) & =
\sum_{\gamma_1,\gamma_2} {{L^0_{\gamma_1}{L^0_{\gamma_2} {\left(L^2_{\gamma_1}+L^2_{\gamma_2}\right)}^{-1/4}} \over {\sqrt{\left|{\det}
(\tilde{M}_{\gamma_1}-I)\right|} \sqrt{\left|{\det}
(\tilde{M}_{\gamma_2}-I)\right|}}}} \left[ {\alpha^{5/4}{\mathcal{A}}E^{1/4} 
\over (2\pi)^{5/2}L_{12}} \cos \left(\Phi_{12} -
{3\pi\over4} \right) \right.  \\
& \left. -{\alpha {\mathcal{L}} \over 16 \pi^2 L^{1/2}_{12}} \cos \left( \Phi_{12} -
{\pi \over 2} \right) + {\alpha^{3/4}E^{-1/4} \over (2\pi)^{3/2}}
\left({{\mathcal{A}} \over 4 \pi L^2_{12}} + {\mathcal{K}} \right) \cos
\left(\Phi_{12}-{\pi\over4}\right) \right],
\end{split}
\end{equation}
\end{widetext}
where $L_{12}=\sqrt{L^2_{\gamma_1}+L^2_{\gamma_2}}$, 
$\sigma_{12}=\sigma_{\gamma_1}+\sigma_{\gamma_2}$ and  
$\Phi_{12}=\sqrt{\alpha E}L_{12}-\sigma_{12}{\pi/2}$.

\end{document}